\documentclass[twocolumn,superscriptaddress,amsmath,amssymb,aps,prb,citeautoscript,10pt]{revtex4-2}
\usepackage{multirow}
\usepackage[utf8]{inputenc}
\usepackage{graphicx}
\usepackage{bm}
\usepackage[colorlinks,allcolors=blue,breaklinks]{hyperref}
\graphicspath{{./figures_tables/}}
\usepackage{color, colortbl}
\usepackage[table, svgnames, dvipsnames]{xcolor}
\usepackage{makecell, cellspace}
\definecolor{Gray}{gray}{0.9}
\usepackage{xcolor}
\edef\restoreparindent{\parindent=\the\parindent\relax}
\usepackage{parskip}
\restoreparindent

\usepackage{titlesec}
\titleformat{\section}{\normalfont\fontsize{11}{11}\bfseries}{}{}{}[]
\titlespacing{\section}{0em}{9pt}{3pt}
\titleformat{\subsection}[runin]{\normalfont\fontsize{10}{10}\bfseries}{}{}{}[.]
\titlespacing{\subsection}{0em}{9pt}{4.4pt}
\usepackage{tikz}

\begin{document}
\renewcommand{\refname}{REFERENCES}
\setcitestyle{super}  
%

\title{Atomic Representations of Local and Global Chemistry in Complex Alloys}
%

\author{M. J. McCarthy}
\email{megmcca@sandia.gov}
\affiliation{Center for Computing Research, Sandia National Laboratories, Albuquerque, New Mexico 87185, USA}
\author{J. Startt}
\email{jstartt@sandia.gov}
\affiliation{Center for Integrated Nanotechnologies, Sandia National Laboratories, Albuquerque, New Mexico 87185, USA}
\author{R. Dingreville}
\email{rdingre@sandia.gov}
\affiliation{Center for Integrated Nanotechnologies, Sandia National Laboratories, Albuquerque, New Mexico 87185, USA}
\author{A. P. Thompson}
\email{athomps@sandia.gov}
\affiliation{Center for Computing Research, Sandia National Laboratories, Albuquerque, New Mexico 87185, USA}
\author{M. A. Wood}
\email{mitwood@sandia.gov}
\affiliation{Center for Computing Research, Sandia National Laboratories, Albuquerque, New Mexico 87185, USA}

\date{\today}

\begin{abstract}
{
The exceptional properties observed in complex concentrated alloys (CCAs) arise from the interplay between crystalline order and chemical disorder at the atomic scale, complicating a unique determination of properties.  
In contrast to conventional alloys, CCA properties emerge as distributions due to varying local chemical environments and the specific scale of measurement.
Currently there are few ways to quantitatively define, track, and compare ‘local' alloy compositions (versus a 'global' label, i.e. equiatomic) contained in a CCA. 
Molecular dynamics is used here to build descriptive metrics that connect a global alloy composition to the diverse local alloy compositions that define it.
A machine-learned interatomic potential for MoNbTaTi is developed and we use these metrics to investigate how property distributions change with excursions in global-local composition space.
Short-range order is examined through the lens of local chemistry for the equiatomic composition, demonstrating stark changes in vacancy formation energy with local chemistry evolution. 
}

\end{abstract}

\flushbottom
\maketitle
\thispagestyle{empty}
\clearpage
\section*{INTRODUCTION}
{
The basic metallurgical idea behind complex concentrated alloys (CCAs) is two-fold: the maximization of the solid solution strengthening effect and phase stabilization via configurational entropy versus enthalpy.
Composed of three or more metallic elements, CCAs represent a family of alloys with a large composition space for property improvement to be explored.
Many physical metallurgy research groups have explored this space and ventured away from the reference equiatomic composition to seek novel compositions that yield improved properties ranging from
high temperature strength~\cite{senkov2015accelerated, yao2016, senkov2018development, coury2019, senkov2019high, startt2022compositional},
to wear resistance~\cite{cui2020wear,du2021mechanical},
to corrosion and oxidation resistance~\cite{liu2019effect, esmaily2020high, scully2020controlling},
to magnetic properties~\cite{zhao2018effect, chaudhary2021accelerated, mishra2021design}.
For instance,~Startt and coworkers~\cite{startt2022compositional} investigated how incremental changes in elemental compositions affected thermomechanical properties in the MoNbTaTi refractory CCA, finding that increases in the concentration of Mo lead to significantly improved yield strength, while higher Ti concentrations lead to greater ductility.
Similarly,~Zhao and colleagues~\cite{zhao2018effect} showed that increasing the amount of Co in a CoCrCuFeMnNi alloy resulted in a higher magnetization saturation moment and an improved soft magnetic response.
}

{
While most studies use only the global alloy composition, $c_{\rm global}$, to characterize and describe an alloy and its properties, the reliance on a single global descriptor can be problematic. 
Though they exist in ordered crystalline states, CCAs typically possess a large degree of intrinsic chemical disorder, leading to large fluctuations in micro- and nanoscale properties.
Those properties can in turn be highly sensitive to local environment and short-range ordering effects.
Such local/global duality adds a new dimension and challenges to the atomic representation of the huge compositional space in these multicomponent alloys.
Atomistic modeling techniques such as density functional theory (DFT) and molecular dynamics (MD) offer ways to quantify the connections between local and global chemistry in these complex alloys.
Recent developments in the training and validation of machine-learned interatomic potentials (ML-IAPs) have made CCAs more accessible in MD, allowing for deeper investigation into local and global property relationships.
Several ML-IAPs have already been developed to study MoNbTaW-based quaternary and quinary refractory BCC alloy systems and have been used to study various CCA properties such as segregation and defect formation~\cite{byggmastar2021modeling}, strengthening mechanisms~\cite{li2020complex}, and dislocation mobility~\cite{yin2021atomistic}. 
In all cases, both the ML-IAP training procedures as well as subsequent analysis were centered around global composition descriptions, potentially obscuring important trends that may arise from local chemical fluctuations. 
These descriptive limitations arise from the lack of a unified language or system to describe variability in chemistry.
}

{
In this work, we are focused on the intricate relationships between the local and global chemistry in complex alloys.
We are particularly interested in the implications these relationships have in estimating alloy properties, validating CCA interatomic potentials' performance through composition space, and in training improved CCA ML-IAPs in the future.
To do this, we begin by defining the composition space of a quaternary CCA (elements labeled A, B, C, and D) in Fig\@.~\ref{fig:fig1}, where the concept of local and global duality is illustrated.
In Fig\@.~\ref{fig:fig1}a, the tetrahedron represents the entire composition space spanned for our quaternary alloy.
The inner volume of this tetrahedron contains all possible quaternary compositions and the center of the tetrahedron represents the equiatomic composition (black star).
The vertices, edges, and faces of the tetrahedron correspond to pure, binary, and ternary compositions, respectively.
From a macroscopic perspective a global composition, $c_{\rm global}$, corresponds to a single point within this tetrahedron for an infinitely large random solid solution structure, such as that in  Fig\@.~\ref{fig:fig1}b.
Conversely, Fig\@.~\ref{fig:fig1}c illustrates the concept of a \emph{local composition}, $c_{\rm local}$,  which can be defined as the concentration of elements found within a spherical volume element of radius $r_{\rm cut}$ centered on any one atom or lattice site, $i$.
Mapping all local compositions for a given $r_{\rm cut}$ from all atoms in a structure yields a point cloud of $c_{\rm local}$ values centered around the parent structure's $c_{\rm global}$, similar to how averaging all atoms' local composition yields $c_{\rm global}$.
Varying $r_{\rm cut}$ changes the number of atoms locally sampled and thus the density of the point cloud, as shown in the two tetrahedra of Fig\@.~\ref{fig:fig1}c.
The variability in $c_{\rm local}$ depends on $r_{\rm cut}$ and will eventually dwindle and converge to $c_{\rm global}$ as the cutoff increases to infinity.
}

{
 \begin{figure*}[h!t]
        \centering
        \includegraphics[width=\textwidth]{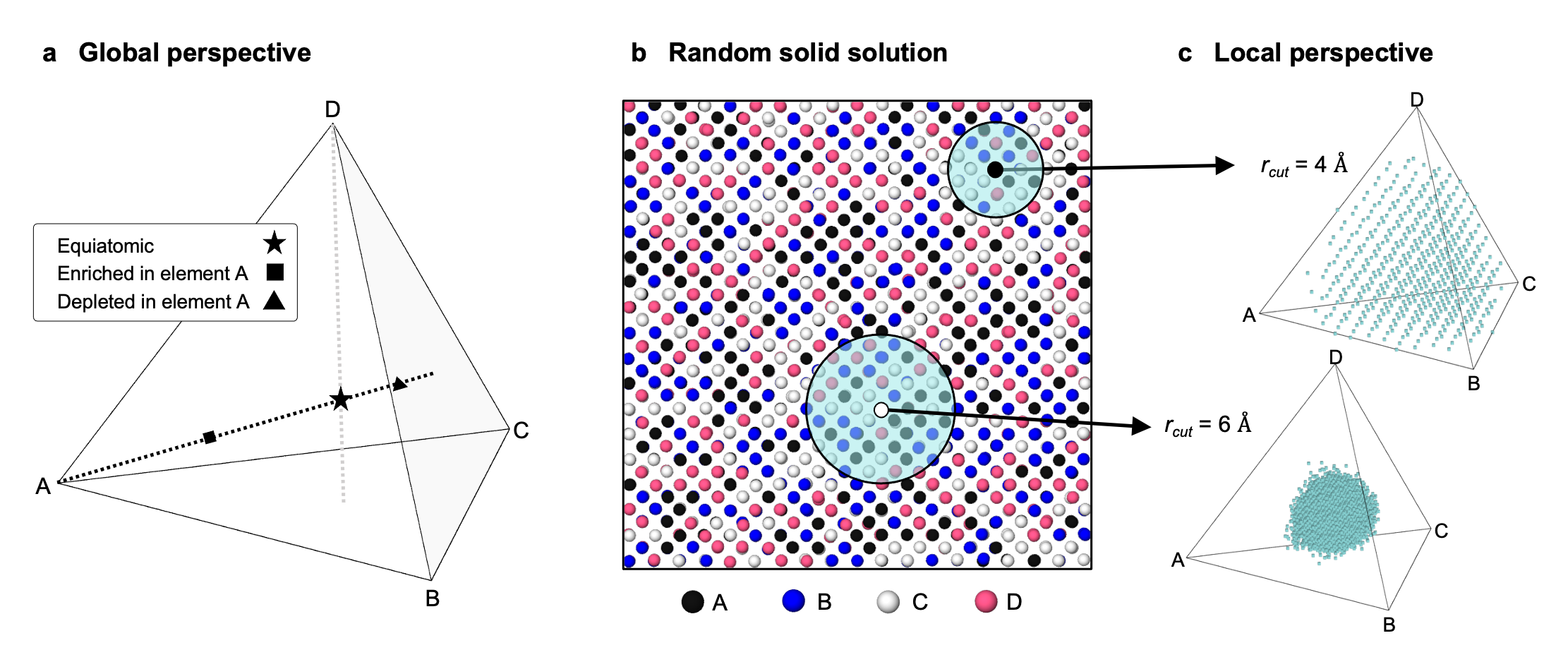}
        \caption{ Illustration of global and local composition spaces for a quaternary alloy of generic elements A B C D. 
        {\bf a}~ Global composition perspective. The location of a quaternary alloy's equiatomic composition in the phase tetrahedron is shown with a star. The vertices, edges, and faces of the tetrahedron correspond to pure, binary, and ternary compositions, respectively. Dotted black line tracks the changes in one element's concentration from pure A at the vertex, to a CCA enriched in A (square), to one depleted of A (triangle) to a ternary alloy consisting of B, C, and D. Gray dotted line starting at element D is included as a guide for the eye.
        {\bf b} An example equiatomic random solid solution structure with elements A, B, C, and D. The mapping of local compositions $c_{\rm local}$ for each atom $i$ for different values of $r_{\rm cut}$ is shown in panel {\bf c}.  
        {\bf c}~ Local composition perspective. An atom $i$'s local composition is calculated by taking the concentration of all elements found within a sampling radius $r_{\rm cut}$ around said atom. Different values of $r_{\rm cut}$, here 4 \AA~(top) vs. 6 \AA~(bottom), capture different numbers of neighboring atoms. Mapping all of a structure's atom's local compositions results in point cloud sizes that depend on $r_{\rm cut}$. 
        }
        \label{fig:fig1}
 \end{figure*}
}

{
The relationship between these two perspectives can be quantified by three characteristic metrics:
a composition deviation, $\lambda$, which measures the distance between two compositions (either global and/or local);
a volume fraction, $V_{\rm f}$, that quantifies the amount of composition space a point cloud spans in the local composition tetrahedron; and
the cutoff radius, $r_{\rm cut}$, which defines a physical scale of interest as well as the chemical resolution tying both perspectives together.
The equations corresponding to each metric are included in the Methods below.

In what follows, we demonstrate how $\lambda$ and $V_{\rm f}$  can be used to draw connections between local and global compositions, and how material properties in CCAs can be affected by deviations in composition at the local scale. 
We start by building an ML-IAP for a four-component CCA using a training database generated from ab-initio calculations. 
The MoNbTaTi alloy is a BCC refractory CCA that has been shown to exhibit not just excellent strength and elastic stiffness but is also prone to significant deviations in strength when incremental changes are made to elemental compositions~\cite{startt2022compositional}. 
Thus, it is hypothesized that local composition fluctuations could prove to have significant effects on material properties and behaviors, making this CCA an ideal choice for this present study. 
We then use this trained ML-IAP to illustrate how to sample a representative local chemistry and the implications of this sampling on scale-dependent property measurement across composition space. 
Finally, we examine how local-global composition sampling changes when incorporating CCA chemical effects such as short-range order.
}
\section*{RESULTS}
 \subsection*{Multi-compositional refractory CCA ML-IAP for MoNbTaTi}
{
In this section, we describe the fitting process and the performance of the MoNbTaTi CCA ML-IAP used in the following sections for analysis. 
The primary target in fitting this potential was to ensure accuracy across a wide range of global composition space, with a particular focus on replicating DFT measured elastic properties. 
Because elasticity is a property measured over an entire simulation cell, and not locally, we did not apply the local composition analysis techniques outlined above to optimize the fitting process. 
Instead, we followed the global composition sampling scheme detailed in the Methods.
For creating fitting data, the global composition was systematically varied via a pseudo-equiatomic composition sampling scheme by adjusting the concentration of one element, labeled \textit{element A}, against the other three elements, labeled \textit{elements B, C, and D}. 

This sampling scheme follows that of the black dotted line in Fig\@.~\ref{fig:fig1}a, illustrating that all global composition training points fall along single lines connecting each vertex (corresponding to a pure element) to its opposing tetrahedral face (corresponding to an equiatomic ternary alloy completely absent of the vertex element). Each DFT training structure was averaged over four special quasirandom structures (SQS) to approximate bulk mixing character.
Starting from the equiatomic composition in the center, each component element A was depleted to 12.5\% at minimum and enriched to 50.0\% at maximum.
}

{
\begin{table}[b!]
\centering
\begin{tabular}{|c|c|c|c|c|}
\hline
\hline
{\bf Composition} & {\bf Property} &  {\bf DFT}~\cite{startt2022compositional} &  {\bf SNAP} & {\bf Error (\%)}\\
\hline
\multirow{6}{8em}{Equi.}
 & $\mathbb{C}_{11}$ [GPa] & 239.6 & 255.3 & 6.6\% \\
 & $\mathbb{C}_{12}$ [GPa] & 129.7 & 139.7 & 7.7\% \\
 & $\mathbb{C}_{44}$ [GPa] & 37.8  & 39.1  & 3.4\% \\
 & $B$ [GPa] & 165.7  & 178.4  & 7.7\% \\
 & $G$ [GPa] & 43.6  & 42.7  & 1.0\% \\
 & $E$ [GPa] & 120.1  & 126.4  & 5.2\% \\
 & $\nu$ & 0.379  & 0.382  & 0.8\% \\
 \hline
 \multirow{6}{8em}{Mo 12.5 at.-\%} 
 & $\mathbb{C}_{11}$ [GPa] & 234.3 & 233.8 & 0.2\%  \\
 & $\mathbb{C}_{12}$ [GPa] & 134.6 & 134.3 & 0.2\% \\
 & $\mathbb{C}_{44}$ [GPa] & 35.6  & 36.0  & 1.0\% \\
 & $B$ [GPa] & 167.8  & 167.5  & 0.2\% \\
 & $G$ [GPa] & 40.8  & 40.1  & 0.5\% \\
 & $E$ [GPa] & 113.2 & 113.6  & 0.4\% \\
 & $\nu$ & 0.388  & 0.387  & 0.2\% \\
\hline
 \multirow{6}{8em}{Mo 50 at.-\%} 
 & $\mathbb{C}_{11}$ [GPa] & 334.4 & 312.5 & 6.5\%  \\
 & $\mathbb{C}_{12}$ [GPa] & 147.3 & 149.3 & 1.3\% \\
 & $\mathbb{C}_{44}$ [GPa] & 50.331  & 50.0  & 0.6\% \\
 & $B$ [GPa] & 209.7  & 203.7  & 2.9\% \\
 & $G$ [GPa] & 64.7  & 60.9  & 5.8\% \\
 & $E$ [GPa] & 175.9 & 166.2  & 5.5\% \\
 & $\nu$ & 0.360  & 0.364  & 1.1\% \\
\hline
\hline
\end{tabular}
\caption{
Elasticity value comparisons from the MoNbTaTi dataset and the SNAP model for the equiatomic random solid solution and an example at Mo enriched to 50 at.\%. Each value was averaged over four DFT special quasirandom structure calculations and five SNAP calculations in MD. 
}
\label{table:tab1}
\end{table}
}

{
For the purposes of this work, we selected one well-performing ML-IAP generated using the above parameters, though many thousand ML-IAPs were generated and tested. 
A schematic of the fitting process is shown in Fig\@.~\ref{fig:fig2}a, consisting of two major parts: a linear regression which connects the DFT data to LAMMPS bispectrum calculations, run in the software FitSNAP~\cite{Rohskopf2023} and a genetic algorithm to optimize SNAP hyperparameters, run in the DAKOTA software~\cite{dakota}. 
More details on each part, and the specifics of fitting to multiple compositions, are included in the Methods. 
Errors on the energies and forces for this ML-IAP are 12 meV and 176 meV/\AA, respectively. 
Representative model errors for elastic constants and moduli for the equiatomic and two Mo compositions (Mo at 12.5 at.\% and 50 at.\%) are shown in Table\@.~\ref{table:tab1}.
All errors for the equiatomic and for a great number of other fitted compositions remain in general below 10\% of their respective DFT values and frequently near or below 5\%.
Fig\@.~\ref{fig:fig2}b plots the bulk modulus B and shear modulus G for a selection of trained compositions for both the trained SNAP ML-IAP (curves) and DFT values (symbols). Values from the trained model's bulk and shear moduli ($B$ and $G$, respectively) are shown in Fig\@.~\ref{fig:fig2}b for 29 selected compositions. Data points from the DFT training (symbols) are averaged over four SQS elasticity calculations (taken from Startt et al.~\cite{startt2022compositional}), and trained SNAP ML-IAP (curves) are averaged over five MD calculations. The error bars of the SNAP curves show the standard deviation over five modulus calculations. Note that, because the DFT and SNAP values for the equiatomic calculations had $< 1$\% error, the symbols have been left off for clarity.
The horizontal axis of these plots follows the concentration of element A, whose species assignment is indicated by colors and symbols. 
In general, the ML-IAP moduli match closely to the expected DFT values for the majority of the trained global alloy compositions.
Depletion or enrichment of Mo, Nb, and Ti from the equiatomic composition are especially well-fit. 
The largest errors are found in alloys with Ta at the highest (approx. 20\% error at 50 at-\% Ta) and lowest concentrations (approx. 8\% error at 12.5 at-\% Ta). 
}

{
\begin{figure*}[h!t]
        \centering
        \includegraphics[width=\textwidth]{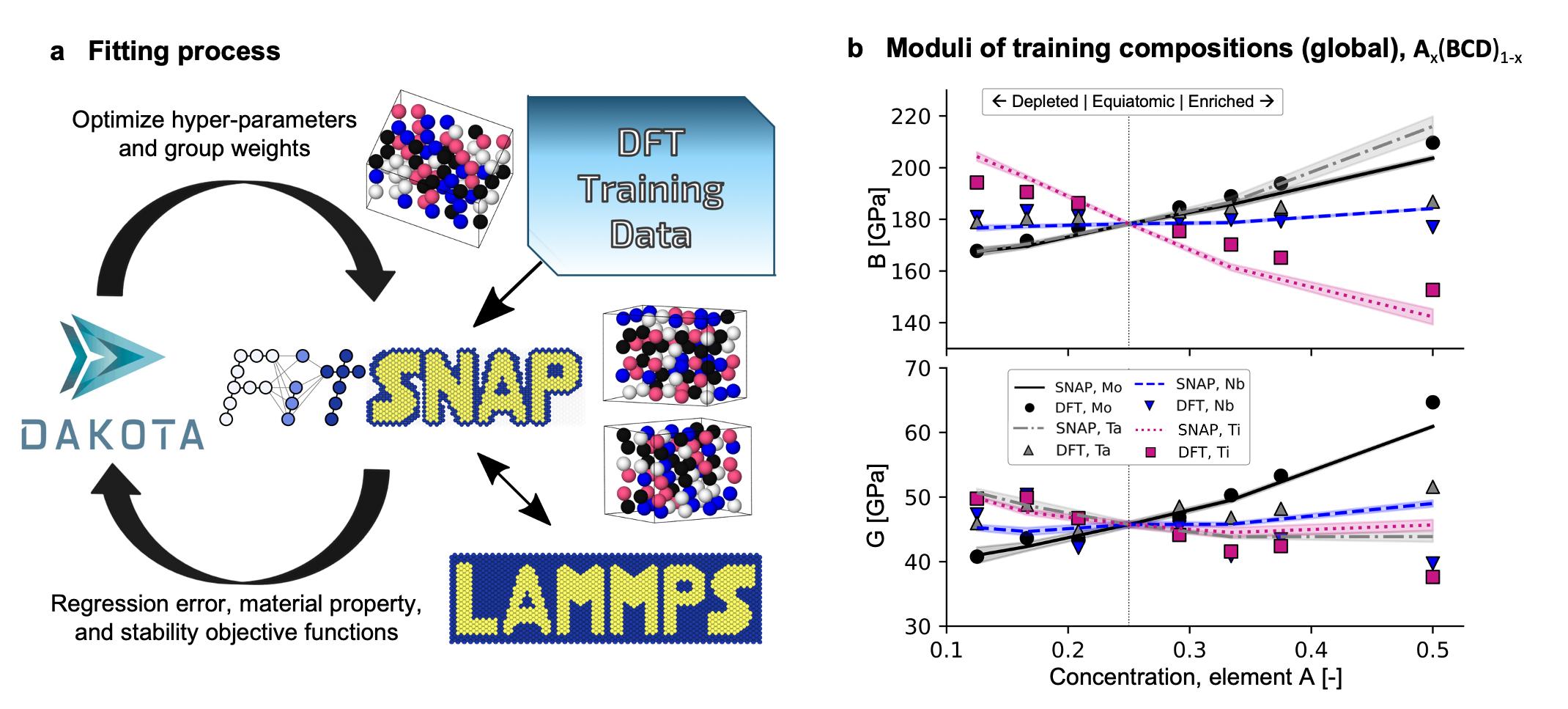}
        \caption{
        {\bf a} Schematic of the ML-IAP fitting process. The FitSNAP software~\cite{Rohskopf2023} performs regression while LAMMPS is used to calculate descriptors. The resulting fit is run through a series of tests, whose errors are used to inform a single-objective genetic algorithm provided by DAKOTA. By iterating thousands of times, the genetic algorithm is able to isolate ML-IAPs with optimal material properties and excellent dynamic stability.
        {\bf b} Calculations of the bulk modulus B (top) and shear modulus G (bottom) for selected compositions using the SNAP MoNbTaTi model. Solid lines follow the SNAP-calculated values for enrichment or depletion of each element in the refractory CCA where corresponding shading indicates the standard deviation. The symbols show averaged data from the DFT training, taken from Startt et al.~\cite{startt2022compositional}.  
      }
        \label{fig:fig2}
 \end{figure*}
}

 \subsection*{Representative Sampling of Chemistry}
{
Given the availability of an IAP, one of the key advantages of MD simulations is their ability to scale. 
By dint of its larger possible size, a single MD simulation cell ($N_{\rm MD} \leq 10^{10}, \leq 1~{\rm \mu m}^{3}$) can sample a vastly wider range of local compositions than even a large ensemble of DFT-sized ones ($N_{\rm DFT}\leq 10^{4}, \leq 10~{\rm \AA}^{3}$). 
This range means that, for a given global composition $c_{\rm global}$ of a random solid solution (\textit{i.e.} Mo$_{\rm x}$Nb$_{\rm y}$Ta$_{\rm z}$Ti$_{\rm (1-x-y-z)}$), a large fraction of $N_{\rm MD}$ will contain many high probability local compositions ($c_{\rm local}$) near $c_{\rm global}$, as well as extreme values of $c_{\rm local}$ that deviate from $c_{\rm global}$. 
The first effort herein is focused on quantitatively defining these length scales where $N_{\rm MD}$ can be considered as a representative volume of some average chemistry, $c_{\rm global}$. 
Caveats to consider are then: would it suffice to examine only the most common environments? 
What represents an uncommon environment?
And how different is it from the most common ones? 
In this section, we illustrate how metrics of distance and volume within composition space, represented by $\lambda$ and $V_{\rm f}$, can be used to answer these and related questions.
}   

{
The first task is to understand the distribution of $c_{\rm local}$ around a single $c_{\rm global}$. 
We define the scalar distance metric $\lambda=|c_{\rm global}$ - $c_{\rm local}|$, taken element wise, that maps any arbitrary local composition's deviation from the global one.
Its maximum value is $\sqrt{2}$ (the distance between any two pure elements using the Euclidian norm), and its minimum value is determined by $r_{\rm cut}$.
The cumulative distribution function (CDF) of $\lambda$ for three different $r_{\rm cut}$ values is shown in Fig\@.~\ref{fig:fig3}a. 
Reflective of the qualitative local perspective data displayed in Fig\@.~\ref{fig:fig1}c, it is clear in Fig\@.~\ref{fig:fig3}a that the median $\lambda$ is inversely proportional to $r_{\rm cut}$.
This makes intuitive sense due to the fact that when fewer atoms are collected in calculating $c_{\rm local}$, the concentration will change more drastically with a single atom swap.
Additionally, these single atom composition changes also explain the more jagged nature of the CDF for smaller $r_{\rm cut}$.
}

{
\begin{figure}[h!t]
        \centering
        \includegraphics[width=0.5\textwidth]{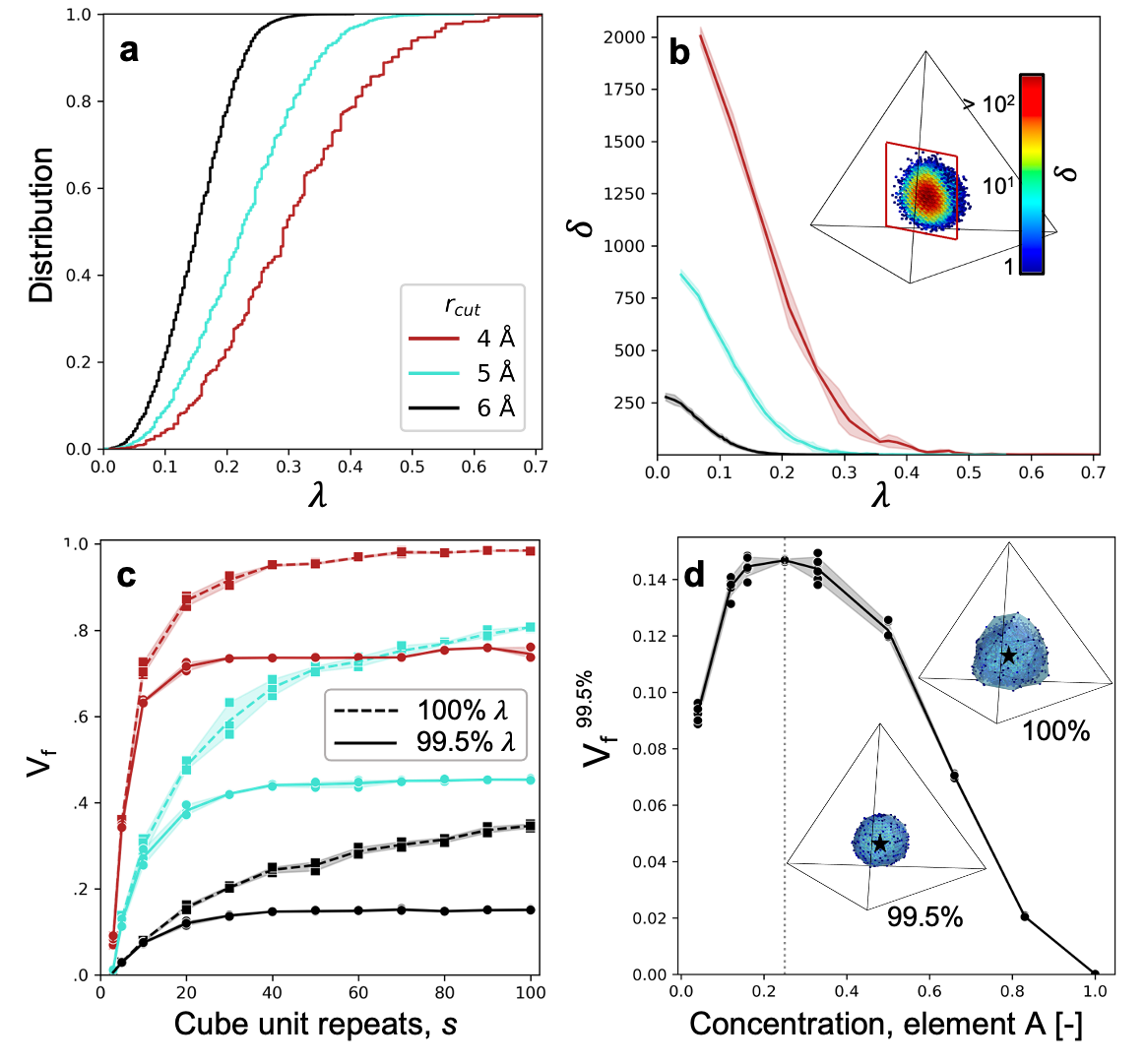}
        \caption{
        {\bf a}~ CDF of $\lambda$ for three values of $r_{\rm cut}$. 
        {\bf b}~ Duplicate local compositions $\delta$ versus the deviation from the global composition $\lambda$ for each $r_{\rm cut}$. Inset is of s = 40 and $r_{\rm cut}$ = 6~\AA~, with the color representing the number of duplicates $\delta$.
        {\bf c}~ Volume fraction $V_{\rm f}$ as a function of simulation cell size, s. Dashed lines show $V_{\rm f}$ calculated with all values of $\lambda$ (as in the top inset of panel {\bf a}), solid lines where 0.5\% of the highest values of $\lambda$ excluded. Excluding 0.5\% of $\lambda$'s largest values not only significantly lowers $V_{\rm f}$, but also stabilizes it to a constant value for all $r_{\rm cut}$ shown.
        {\bf d}~ Volume fraction $V_{\rm f}=V_{\rm f}^{99.5\%}$ for a range of compositions included in the training set (see Fig\@.~\ref{fig:fig1}a for reference), using $s=40a_0$ and $r_{\rm cut}$ = 6~\AA~ averaged over 5 runs for one example element. Insets show the convex hull geometry of $r_{\rm cut}$ = 6~\AA~ with all $\lambda$ included (top), and with the largest 0.5\% $\lambda$ excluded (bottom).
        }
        \label{fig:fig3}
 \end{figure}
}

{
From the discreteness in single atom changes to $c_{\rm local}$ and on-lattice measurement of $\lambda$ comes another metric, the \textit{duplicate local compositions}, $\delta$, which is a count of the number of atomic environments that have identical local composition values. 
Fig\@.~\ref{fig:fig3}b shows $\delta$ as a function of $\lambda$ for different values of $r_{\rm cut}$. 
Projected into the composition tetrahedron, the inset shows local composition data for $r_{\rm cut}$ = 6~\AA~with a log-scaled color bar. 
For $r_{\rm cut}$ = 4~\AA, the smallest possible compositional change is $\sim$5.77 at-\%, where this definition of length scale corresponds to the two nearest-neighbor shells in BCC. 
The smaller cutoff radius results in higher degeneracy of observed $c_{\rm local}$, and coarser $\lambda$ spacing.
Though duplicate compositions are chemically degenerate, they are likely to be structurally unique, just as is true for global compositions as a whole. 
The differences between each duplicate environments' atomic arrangements gives rise to a distribution of atomic energies (as predicted by the SNAP ML-IAP developed herein) per $c_{\rm local}$.
}

{
Due to the coarse chemical representation of short cutoff distances, the volume contained by a bounding surface of $c_{\rm local}$, $V_{\rm f}$, shown in Fig\@.~\ref{fig:fig3}c, also increases.
As the occurrence of extrema local chemistries is more probable at short cutoff distances, these outlier points are filtered out by a modest $0.5\%$ rejection of the highest $\lambda$ values (Fig\@.~\ref{fig:fig3}c, solid lines).
Compositional volume fraction $V_{\rm f}$ data are plotted against the size of a cubic simulation cell, where the number of unit cell replicas, $s$, spans atom counts capable of being represented in DFT, 54 ($s=3$), to those only possible in MD, $\sim 1\cdot10^6$ ($s \geq 70$).
Sampling of $\lambda$ is taken from 5 unique random solid solution cubes per data point with unit side length of $s$.
Shaded bands around each line and point encompass the standard deviation for each $r_{\rm cut}$.
}

{
At small cell sizes, before $V_{\rm f}$ has been saturated, not enough unique $c_{\rm local}$ configurations will be present for the simulation cell to be chemically representative of $c_{\rm global}$.
The apparent lack of saturation in the data for $r_{\rm cut}$ = 4~\AA~is mathematically explained by the probability for $c_{\rm local}$ to be a single element species. 
Element labels drawn at random for the 15 nearest atoms equal to the central atom type is $p=(\frac{1}{4})^{15}$, which multiplied by the $10^7$ atom environments sampled for $s=70$ is not a vanishingly small probability($\approx 0.93\%$).
The same random draw where $r_{\rm cut}$ = 6~\AA~can be safely assumed as a complete chemically representative volume element.
Filtering out the largest 0.5\% of $\lambda$ values ($V_{\rm f}^{99.5\%}$) leads to a significant drop in volume fraction, as shown in solid lines in Fig\@.~\ref{fig:fig3}c. 
Fig\@.~\ref{fig:fig3}d captures saturated $V_{\rm f}^{99.5\%}$ values while moving along a tie line of one element A (e.g., where A = Mo, then  ${\rm Mo}_{\rm x}({\rm NbTaTi})_{\rm 1-x}$). 
There it can be seen that the chemically representative volume element is more compact for compositions near a dilute single component or ternary composition of the CCA. 
For example, the chemically representative volume element for random solid solution equiatomic cubes sampled with $r_{\rm cut}$ = 6~\AA~is $V_{\rm f}^{99.5\%}=0.15$ located directly in the center of the phase tetrahedron. 
In contrast, chemically representative volume elements for near-ternary alloys (where the concentration of element A approaches $<$ 10 at.-\% of element A) or element A-enriched alloys (increasing at-\% of element A) are comparatively smaller than the equiatomic. 
The same measure taken for a $c_{\rm global}$ near a single element vertex would result in a comparatively small $V_{\rm f}^{99.5\%}$, given the fewer number of possible local chemical arrangements.
It is also important to note that, though we have chosen to filter the largest 0.5\% values of $\lambda$, one could also choose heavier filtering, which would decrease $V_{\rm f}$ proportionally.
As a visual guide, the two inset images of Fig\@.~\ref{fig:fig3}d show two cases of the bounding surface formed by each of the $c_{\rm local}$ values, either where the full CDF (top inset) is used, resulting in a rough surface, or when the lowest 99.5\% values of $\lambda$ (bottom inset) is used, yielding a smooth surface. 
The exact choice of a saturation filter is less important than having the ability to make informed decisions about the degree of chemical accuracy desired for CCA modeling and simulation.
}

{
Taking this quantitative measure one step further, the \textit{complexity} of an alloy can be defined by its $V_{\rm f}^{99.5\%}$, which has merit for alloys of both fewer and more elements included in them. 
While the visualization of the volume contained by the collection of $c_{\rm local}$ in five or more element composition space is challenging, using these relatively simple measures of complexity make comparisons possible. 
As will be explored in an upcoming section, complexity metrics also enable an assessment of the degree of chemical change within a single $c_{\rm global}$.
For example, where short-range order is present in a given CCA, $\lambda$, $\delta$ and $V_{\rm f}$ will reflect the unique evolution with respect to an equivalent random solid solution at the same $c_{\rm global}$.
Furthermore, when selecting a computational method for predicting properties of CCAs, these chemically representative volume element measures confirm that DFT-sized simulation cells, which have maximal atom counts on the order of $10^2$ ($s\leq 5$) for a single cell, sample far too small a fraction of the chemically representative volume element (which includes all duplicates per local composition $\delta$) to capture global CCA properties accurately. 
}

{
Having determined a reasonable simulation cell size from which to properly sample CCA properties (i.e., $s \simeq 40$), we now turn predictions of material properties that can be reasonably assumed to be of the same length scale as these $r_{\rm cut}$ definitions.
}

\subsection*{Scale-Dependent Property Measurement}
{
Material properties in CCAs are generally expressed as distributions rather than as single, averaged values typical of chemically simpler systems.
The question then becomes whether this property distribution is explained by the $c_{\rm local}$ distribution defined herein.
This assertion is critically dependent on the characteristic length scale of the property of interest, and whether that length scale demonstrates the chemical variability discussed previously.
Example properties and their associated length scales could include: elastic moduli that encompass the entire chemically representative volume element, dislocations that sample as many $c_{\rm local}$ as atoms along the line, or point defects that may (depending on the strain field generated) only reflect a single $c_{\rm local}$.
In order to enable high-throughput alloy design in CCAs, we need an ability to characterize which properties are sensitive to the distribution of $c_{\rm local}$, or can be assigned to $c_{\rm global}$. 
A material defect whose properties are highly spatially localized is the vacancy formation energy. 
It has been shown that vacancy formation and migration energies in CCAs have varied distributions~\cite{zhou2022vacancy} which can for example affect the evolution of radiation damage cascades~\cite{cusentino2020compositional,li2019first}.
}

{
Utilizing the SNAP ML-IAP described above and in the Methods, we sample the vacancy formation energy for each element species, measuring $\lambda$ in order to address how properties are distributed around a given $c_{\rm global}$.
The local chemical definition coincides with the interaction range of the ML-IAP where $r_{\rm cut}$ = 6~\AA.
This distance sufficiently resolves the strain field at a vacancy, where the outermost (see Fig\@.~\ref{fig:fig1}b) neighbor shell shows local strain values less than $6\cdot10^{-4}$ as measured by tools made available in OVITO~\cite{stukowski2012elastic}.
}

{
Fig\@.~\ref{fig:fig4} shows the vacancy formation energy, $E_{\rm vac}$, of removed Nb atoms versus $\lambda$, with colors indicating CCA compositions that are Nb-depleted and Nb-enriched in panels a and b, respectively. 
Details on vacancy formation energy calculations can be found in Methods.
The contour plots indicate regions of similar data point density, calculated using a kernel density function. 
The outermost contour contains 90\% of data points, with each line thereafter delineating a 20\% change in point density up to the innermost line, which encompasses the final 1\% of data.
This smallest contour indicates the most probable values from the $\lambda$ and $E_{\rm vac}$ distributions.
The univariate distribution plots provide another means of interpreting point densities with respect to its parallel axis. 
We keep the equiatomic composition included in both plots as the aforementioned baseline CCA distribution example. 
}

{
 \begin{figure*}[h!t!]
        \centering
        \includegraphics[width=\textwidth]{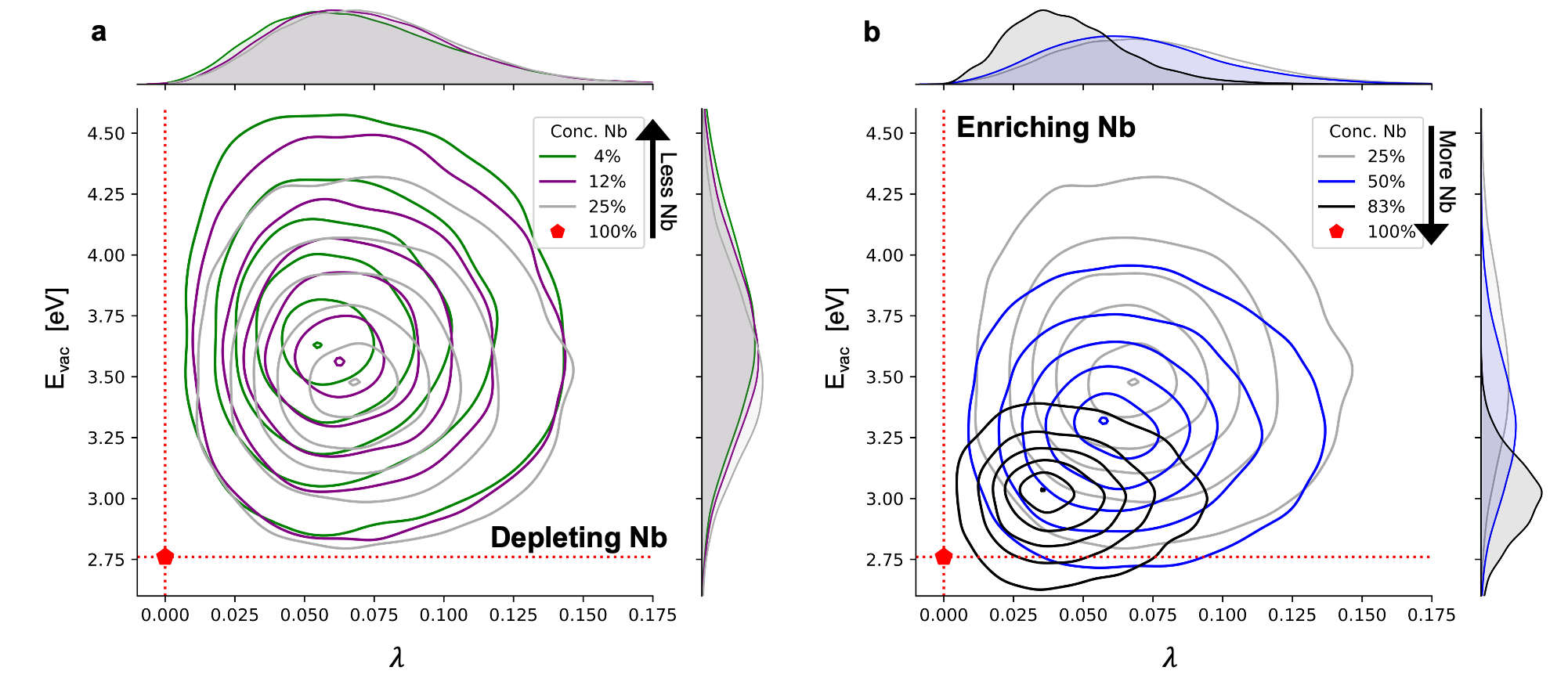}
        \caption{
        {\bf a-b} Vacancy energy  trends vs. $\lambda$, the deviation from a chosen global composition. Each plot features distributions for Nb vacancies. Panel {\bf a} illustrates trends while depleting Nb and panel {\bf b} shows those for enriching Nb in the CCA, with each global composition indicated by different colors. Contour lines, calculated using a kernel density estimate, enclose (outer to inner) 90\%, 70\%, 50\%, 30\%, and 10\% of data points, and the innermost circle the most probable 1\%. The dotted red lines show the average vacancy energy in a pure Nb sample, which corresponds to $\lambda = 0$. The X- and Y-axes show the univariate distribution for the vacancy energies and deviations from global, respectively.
        }
        \label{fig:fig4}
 \end{figure*}

}

{
In Fig\@.~\ref{fig:fig4}a, the Nb depleted compositions, which moves toward the MoTiTa ternary face in the composition tetrahedron, show very little change in this localized property with respect to the composition deviation and maintain a $E_{\rm vac}$ as seen in the average equiatomic CCA (= 3.40 eV) than pure Nb (= 2.76 eV). 
The converse is true for the enriched side of the tie line.
The composition deviations gradually decrease in tandem with the property measurement, converging on the pure element's value.
}

{
Given the locality (as determined by the rapidly diminishing strain field) of a vacancy, it is reasonable to attribute the formation energy to the structure change within the range of the $c_{\rm local}$ definition.
Therefore, it is incorrect to assign $E_{\rm vac}$ to $c_{\rm global}$.
Rather, each $c_{\rm local}$ carries its' own property value and the global composition only determines the range of compositions where that property is averaged based on the chemically representative volume element.
If a larger $r_{\rm cut}$ were used to define the property measurement (and thus the local composition) it would have the effect in a random solid solution to concentrate $c_{\rm local}$ to smaller values of $\lambda$.
But this effect is subverted by allowing for short-range order to develop.
In a chemically-ordered CCA structure, more exotic local environments will be expressed in comparison to a random solid solution.
As will be shown in the following section, these environments give even more merit to assigning properties to local chemistries than global averaged values. 
}

\subsection*{Local composition and short-range order}

{
As useful as a \emph{truly random} solid solution is for general modeling purposes, it is unlikely to accurately reflect the equilibrium state of real CCAs.
Far more probable is the development of at least some degree of short-range order \cite{ferrari2020frontiers,antillon2021,miracle2017critical}.
Though small in CCAs by design, finite differences in chemical potential between species imply that  some reordering is highly probable, especially given factors such as intrinsic defects and thermo-mechanical processing steps that can activate such reordering.
In terms of the metrics defined above, even mild degrees of short-range order evolution will significantly change the geometry of the local composition point cloud.
The values of $\lambda$ that are chemically representative,  $V_f$, $\delta$, and what constitutes an outlier will shift in turn.
Critically, these all will change even for a single global composition label.
This means that, for CCAs, global composition not only obscures important scale-dependent property measurements for the random solid solution, but also lack descriptive power in capturing shifts in properties due to chemical interactions.
In this final section, we will use the above-defined metrics to explore how short-range order alters both local composition space and CCA properties.
This portion of the study is critically dependent on the availability of an accurate energy model, even for extrapolative local chemistries.
}

{
For the MoNbTaTi system specifically, hybrid Monte Carlo MD (MC/MD) and DFT studies of CCAs containing Mo, Ta, and Nb show that Mo and Ta strongly favor each other as neighbors and can form B2 phases~\cite{widom2014hybrid,li2020complex,byggmastar2021modeling,kormann2017long}. 
Hybrid MC/MD simulations of the SNAP MoNbTaTi featured in this work show exactly the same trend, which is clearly reflected in changes in the local composition point cloud as compared to an initially randomly-ordered sample.
Details of the setup of the hybrid MC/MD simulations are included in the Methods.
Fig\@.~\ref{fig:fig5} demonstrates how chemical reordering in an equiatomic sample is reflected in local composition space through time, using $r_{\rm cut}$ = 6 \AA~ and a simulation cell with 128,000 atoms ($s=40$). 
Initially, the structure begins as a random solid solution (Fig\@.~\ref{fig:fig5}a), reflected in the spherical shape of the local composition point cloud.
Note that the colors of the points reflect the species of the central atom being sampled.
Fig\@.~\ref{fig:fig5}b tracks the evolution of both $V_{f}^{99.5\%}$ and the cell's Warren-Cowley parameters~\cite{cowley1950approximate} for the first nearest-neighbor shell $\alpha_1$.
Near-zero values of alpha indicate random ordering, negative values indicate species pair attraction, and positive values indicate repulsion.
Further details of the calculation of $\alpha_1$ are described in the Methods.
By the final simulation step at $t=150$ ps, shown in Fig\@.~\ref{fig:fig5}c, the structure has become clearly ordered.
This change in local chemistry is reflected in the point cloud geometry, which has narrowed and spread into new local composition regions.
The volume fraction has significantly increased, from an initial $V_{f}^{99.5\%}=0.15$ to approx. 0.41 in the chemically-ordered cell.  
Especially notable is a shift towards the binary Mo-Ta phase tie line, demonstrating these elements' capacity to form B2 phases in CCAs.

The favored swapping towards Mo-Ta pairings deplete equiatomic local compositions and shift those points into Ti- and Nb-richer regions of the tetrahedron.
Both of these trends correspond to the pairs with the largest negative values of $\alpha_1$ at $t=150$ ps, namely Mo-Ta, Ti-Ti, and Nb-Nb.
Supplemental Fig\@.~1 shows the evolution of average potential energy through time, as well as a visualization of B2 ordering.
}

{
The structural changes arising from short-range ordering will leave their mark on the distribution of local chemical properties relative to the random solid solution.
Fig\@.~\ref{fig:fig6} illustrates how evolved short-range order alters the vacancy formation energy distributions of the initially-random equiatomic cell.
In both panels, dashed lines indicate the initial states of vacancy energies, and solid lines show the result after reordering.
All elements except Ti exhibit significant shifts in vacancy formation energies, as shown in Fig\@.~\ref{fig:fig6}a. 
Mo atoms undergo clear changes in the shape of the property distribution, with a total cumulative change of -0.28 eV especially represented in the lowest 50\% of the curve.
Ta undergoes an overall increase of 0.61 eV in formation energies. 
All Nb vacancy formation energies decrease in the ordered state by -0.50 eV.
The highest vacancy energies of Ti increase by 0.08 eV, the smallest change of all elements.
Negative (favorable) vacancy formation energies when removing Ti are reflective of the metastability of Ti in a BCC phase.
The unusual shift in the energy distributions of Mo vacancies is explored in more detail in Fig\@.~\ref{fig:fig6}b, with respect to the composition deviation $\lambda$.
While there is some overlap in the contours of the random and ordered structures, the energy values have split into two distinct regions at larger values of $\lambda$, one centered near $E_{vac} \approx 3.25 eV$ and the other much lower at $E_{vac} \approx 1.25 eV$.
An analysis of the characteristic local compositions at $\lambda > 0.3$ indicates that Mo vacancies in the lower energy region have local compositions especially rich in Ti-Nb.
The higher-energy region, which generally also overlaps the random structure's energies, is characterized by Mo-Ta enriched local environments.
}

{
Though further detailed analysis of trends in $E_{vac}$ is beyond the scope of the current work, the additional information given by $\lambda$ enables straightforward segmentation of trends in atomic-scale properties.
These results illustrate the potential power of local composition analysis, highlighting other major benefits of MD studies and ML-IAPs for CCAs.
For example, the local composition analysis techniques developed here can be used to uncover the degree of short-range order change and shift in chemical space from the initial $c_{\rm global}$ not only from two single points in time, but also throughout an entire MC/MD trajectory.
Such information could be used to correlate degrees of short-range ordering with e.g. information gathered from advanced microscopy data \cite{zhang2020short}, creating a potential bridge between the timescales of experimental and computational ordering measures.
The above examples are viewed as avenues for future development of these alloy chemical complexity metrics.
}

{
 \begin{figure*}[h!t!]
        \centering
        \includegraphics[width=\textwidth]{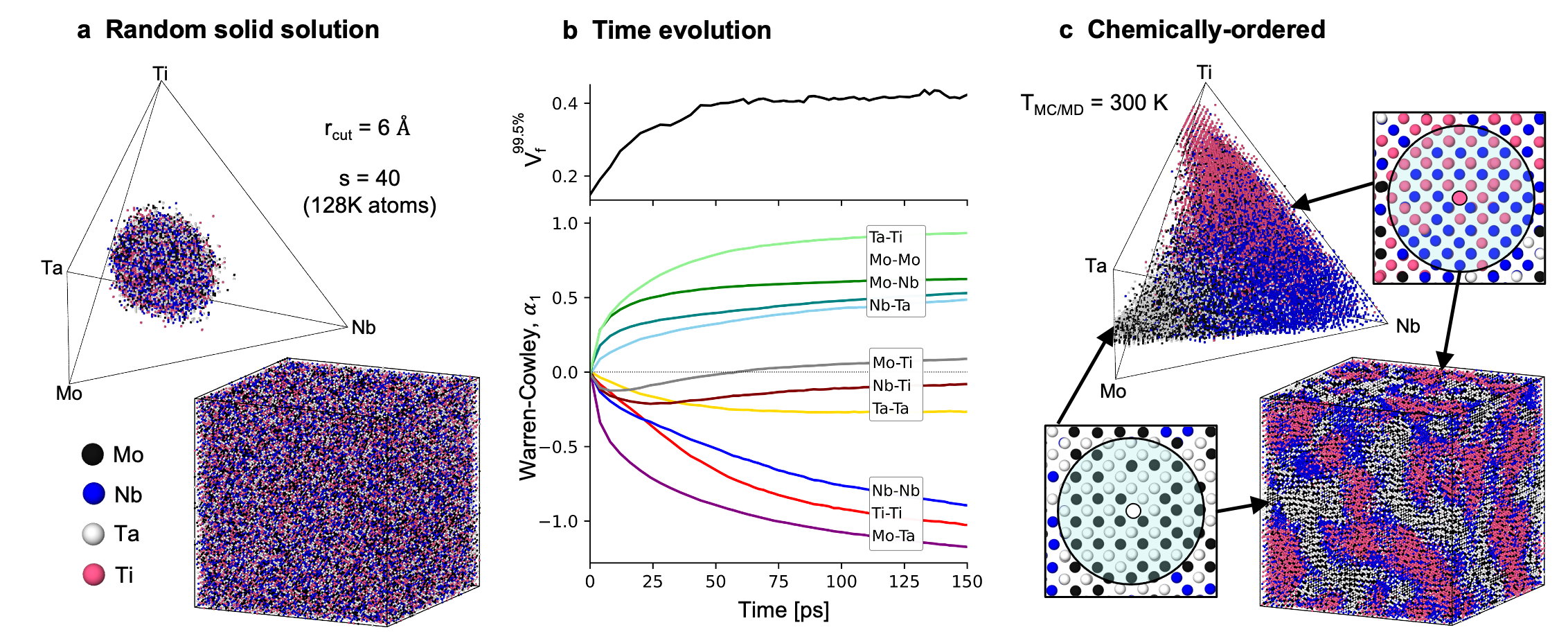}

        \caption{
        Evolution of the local composition tetrahedra ($r_{\rm cut}$ = 6 \AA~) and physical samples of a 128,000-atom equiatomic simulation of hybrid Monte Carlo/molecular dynamics at $T_{MD}$ = $T_{MC}$ = 300 K for $t=150$ ps (768,000 MC atom swap attempts). 
        The colors of the points in the tetrahedra correspond to the species of the sampled central atom.
        {\bf a} The state of the composition tetrahedron at $t=0$ ps. The local composition point cloud is spherical, reflecting the initially random atomic ordering. 
        {\bf b} Change of the volume fraction $V_{f}^{99.5\%}$ (top panel) and the Warren-Cowley parameter for the first nearest-neighbor shell (bottom panel) through simulation time. Negative numbers indicate pair attraction and positive indicate pair repulsion. The most attractive pair by this criteria is Mo-Ta, followed by Ti-Ti and Nb-Nb. 
        {\bf c} The initially-spherical composition point cloud spreads out into broader regions of the local composition tetrahedron, saturating parts of the binary Mo-Ta phase tie line, centered around 50 at.\% and expanding into CCA compositions roughly spanning the entire Ti-Nb tie line.
        }
        \label{fig:fig5}
 \end{figure*}
}

{
 \begin{figure}[h!b!]
        \centering
        \includegraphics[width=0.5\textwidth]{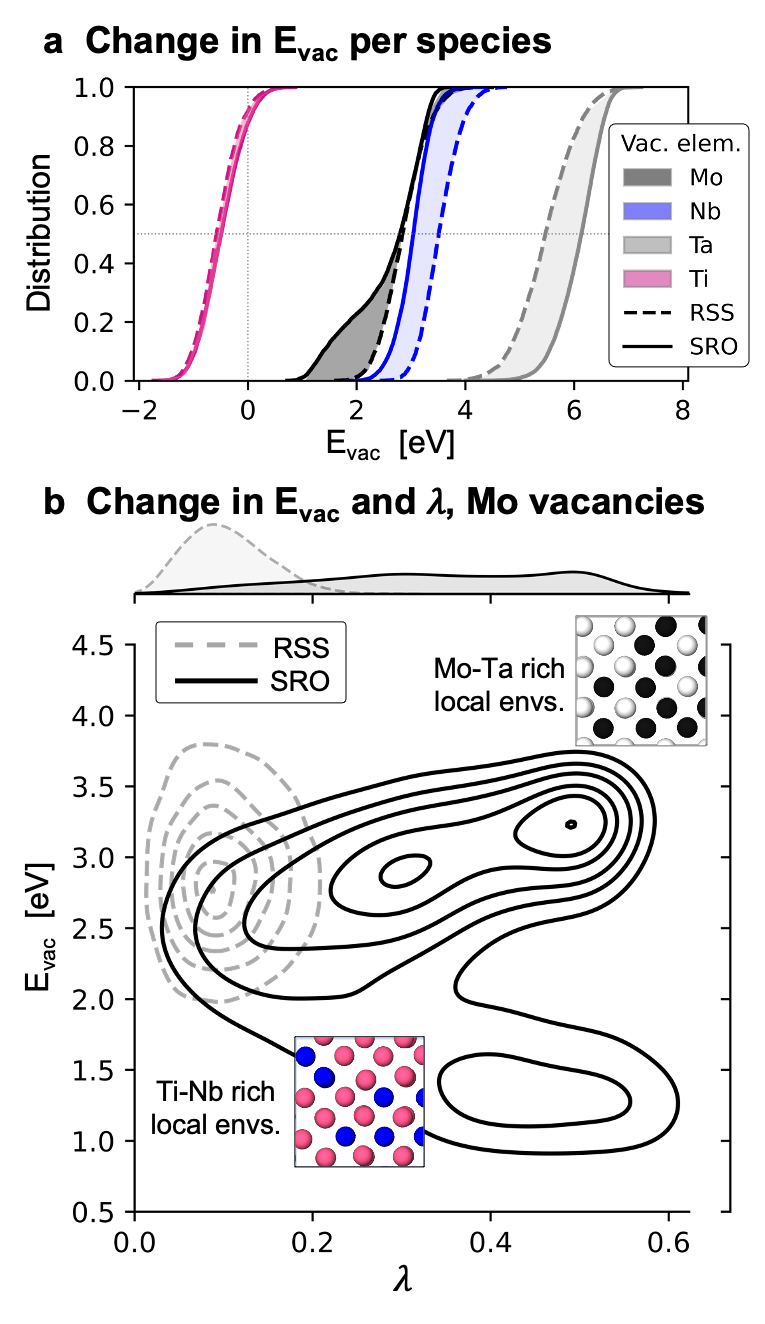}

        \caption{ 
        {\bf a} Cumulative distribution functions of the vacancy formation energies $E_{\rm vac}$ for each element in the random solid solution (RSS, dashed curves) and short-range ordered (SRO, solid curves) cells from Fig\@.~\ref{fig:fig5}, panels {\bf a} and {\bf c} respectively. The shaded regions indicate the shift in energies due to chemical ordering.
        {\bf b} $E_{\rm vac}$ and $\lambda$ for Mo vacancies from panel {\bf a}. The contour outlines share the same parameters as those from Fig\@.~\ref{fig:fig4}. Two regions of different average energies can be observed at values of $\lambda \gtrsim 0.3$, which can be further analyzed to isolate local environmental trends. The region with distinctly lower energies ($\approx 1.25 eV$) is correlated with local compositions rich in Ti-Nb.
        }
        \label{fig:fig6}
 \end{figure}
}

\section*{DISCUSSION}
{
While the details of the model development were not the focus of this work, generating an interatomic potential for a CCA has brought about unique perspective on measuring material properties in CCAs. 
What makes classical molecular dynamics a powerful tool to study materials is the computational efficiency derived from the key approximation that interactions between atoms are localized within some distance.
In turn, we have addressed how this locality has derived alloy complexity measures such as the chemically representative volume element and volume occupied in composition space for a given global composition.
Additional scrutiny is now required on our computational methods when predicting properties of CCAs due to these measures of chemical complexity and locality of a property, which we believe will cause a dramatic shift in this field.
}

{
Looking forward, data of $c_{\rm local}$ and chemically representative volume elements of CCAs should be universally used in training ML-IAPs themselves to ensure proper chemical fidelity. 
Clear metrics of chemical complexity allow confirmation that targeted regions of composition space have been adequately represented in training, and enables the automated generation of new training structures where alloy chemistry data is lacking.
Looking beyond model validation, these techniques could also open new possibilities in scale bridging.
In principle, one could compute material properties in a pre-defined grid in composition space.
When querying a new alloy of a desired (or measured) $c_{\rm global}$ a property can then be inferred based on averaging nearby $c_{\rm local}$ and observing whether those points are contained within the surface of a chemically representative volume element.
This type of prediction would allow MD results to inform property variability at much larger (micro-, meso-) scales as well because chemical inhomogeneities can be defined at numerous length scales. 
For example, if experiments or CALPHAD indicate that certain phases or regions of global composition should be present in CCAs, the chemically representative volume elements for those phases can be added to the training process to ensure models are sampling chemical environments (and also material properties) adequately.
This kind of \textit{diverse-by-design chemical training} approach could be expanded to allow CCA models to capture important phenomena not only through composition space, but also temperature and pressure.
Such capabilities would greatly expand MD simulation utility in becoming truly integrated into workflows of alloy design and discovery.
}

{
All levels of property analysis involve choices about length scale - indeed, navigating those choices is perhaps the most fundamental challenge of materials science. 
Thus, what is needed to bring modeling and simulation efforts in closer concert with experimental capability is a review of the definition of critical length scales.
When practitioners of \textit{ab initio} methods identify a material with its global composition, it should be cast as one of many local composition measurements in MD.
The same can be said for global composition in MD - it is a very local measurement when compared to even the most resolved experimental measurements in, for example, atom probe tomography.
If we are to achieve high-throughput, or even a baseline capability to screen for optimal CCA compositions, these ideas of chemical representation of length scales need to be addressed more completely.
The results herein bring quantitative measures to this challenge, and address how scale-bridging efforts between DFT and MD efforts can resolve the atomic length scale of chemical representation and its influence on simple material properties.
}

\section*{METHODS}

\subsection*{Definition of local and global characteristic metrics}
{
For a simulation cell of arbitrary size, a single point in global composition space $c_{\rm global}$ is equivalent to the average chemistry of all its local atomic environments, $\bar{c}_{\rm local}$:

\begin{equation}
c_{\rm global} = \frac{1}{N}\sum^{N}_{i=1}c^{{\rm atom}, i}_{\rm local} =
\bar{c}_{\rm local}(r_{\rm cut}) 
\label{avg_comp}
\end{equation}

where $N$ is the total number of atoms (i.e., local chemical environments), $c^{{\rm atom}, i}_{\rm local}(r_{\rm cut})$ is the local composition of atom $i$ given a sampling radius of $r_{\rm cut}$. 
This metric considers the joint probability of chemical occupations of all neighbor shells within $r_{\rm cut}$, though for smaller $r_{\rm cut}$ can be approximated by other well-known chemical ordering approaches (e.g., an average of Warren-Cowley parameters~\cite{cowley1950approximate}). 
}

{
For the analysis of local composition, we define two metrics in composition space, the \textit{composition deviation} $\lambda$, and the \textit{convex hull volume fraction} $V_{\rm f}$. The former value is the magnitude of a vector mapping of any two points in composition space:

\begin{equation}
\lambda^{2} = \sum_{i = \\{\rm Mo,Nb,Ta,Ti}}(c_{\rm global}^{i}-c_{\rm local}^{i})^{2}
\label{lambda}
\end{equation}

The latter value requires an extra step. We can create a surface using the outermost points in the point clouds shown in \ref{fig:fig1}b, or in other words, a convex hull. The volume enclosed by the surface of that convex hull $V_{hull}$ encompasses a certain space in a unit tetrahedron, whose side length in 4 dimensions is $\sqrt{2}$. By dividing the hull volume by the volume of a unit tetrahedron, we gain a fraction representing the span of a given sample in composition space:

\begin{equation}
V_{\rm f} = V_{\rm hull}\cdot\frac{1}{3}
\label{eq:Vf}
\end{equation}
}

\subsection*{Density functional theory calculations for training set} 
{
DFT training data for the MoNbTaTi ML-IAP developed in this work is based on the work of Startt et al.~\cite{startt2022compositional}. 
All data for that study was generated using the Vienna Ab-Initio Simulation Package (VASP)~\cite{Kresse1993, Kresse1994, Kresse1995} using plane wave basis sets for the orbital wavefunctions and the the projector augmented wave (PAW) method~\cite{Blochl1994, Kresse1999} to describe interactions between the core and valence-state electrons.
The CCA training data in this work was constructed around two primary composition classes: compositionally varied quaternary compounds and pure elemental systems.
Within each class, DFT calculations were performed to extract forces from several material state conditions to be used in the training of the ML-IAP. These states included: structures with displaced atoms, isotropically and uniaxially strained structures, surfaces, and high-temperature NVT ensemble states. 
}

{
Regarding the simulations for the quaternary compounds, we calculated the aforementioned properties for specific compositional ranges following the composition nomenclature: A$_{\rm x}$(BCD)$_{\rm 1-x}$, where all four metals (Mo, Nb, Ta, and Ti) were taken in turn as the 'A' component.
The value of x was sampled at x = 0.125, 0.1667, 0.2083, 0.25, 0.2917, 0.3333, 0.375, and 0.5 atomic \%.
For quaternary compounds, we performed structural relaxations of four unique special quasirandom structure (SQS)~\cite{Vandewall2013}, 72-atom supercells (a $4 \times 3\times 3$ multiplication of the 2-atom conventional BCC unit cell) at each composition point.
These structures were modelled using a $3 \times 5\times 5$ Monkhorst-Pack k-point grid.
For all four structures belonging to each composition, an atomic displacement calculation (\textit{i.e.} an elastic constant calculation) was performed, where each atom was displaced a total of six times (two times $\pm$ along each Cartesian coordinate). 
For each displacement all atoms are fixed in place and the electronic wave-function and atomic forces are minimized.
For other material conditions, a subset of calculations were performed over a reduced range of composition for x = 0.1667, 0.25, 0.3333 atomic \%.
Uniaxially and isotropically strained supercells were modelled for this reduced composition range and over compression to elongation ranges of 0.86 to 1.06 and 0.94 to 1.03 times the equilibrium lattice spacing, respectfully. 
Surface supercells were modeled for the BCC $100$ (72 atoms), $110$ (72), $111$ (162); and $112$ (128) surfaces. 
Lastly, thermodynamic ab-initio MD (AIMD) NVT simulations were performed at temperatures of 300K, 1200K, 2400K, and 3200K, and at volumes set to match the expected thermal expansion as predicted by~\cite{startt2022compositional} at each temperature.
}

{
Regarding simulations for the pure elements, we simulated each elemental component of the MoNbTaTi quaternary individually in their pure ground state structures (\textit{i.e.,} Mo, Nb, Ta - BCC, Ti - HCP (hexagonally close-packed)).
These systems were modelled over a similar set of deformation, strain, and thermodynamic conditions as the quaternary compounds, with some notable differences.
Given the simplicity of the atomic structures, structural relaxations were performed on smaller simulation cells, utilizing far denser k-point grids.
Additionally, surface slabs were constructed and minimized for the 8-9 most stable surfaces known for each element as listed on the materials project website~\cite{Anubhav2013}.
AIMD NVT simulations at were carried out at 300K, 1200K and one point at least 100K above the known melting temperature of each metal. 
Volumes at each temperature were again set to match the expected lattice constant according to known thermal expansions.
}

{
For both classes of simulations, certain simulation parameters were kept constant.
Notably, plane-wave energy cutoffs were kept at 400 eV for all material systems. 
Wavefunction energy and ionic force minimization criteria were set to 10$^{-6}$ eV and 0.02 eV/\AA$^{2}$, respectively. 
Partial orbital occupancies were set according to a Gaussian smearing scheme using a smearing width of 0.02 eV. 
Exchange and correlation effects were handled according the commonly employed Perdew, Burke, and Ernzerhof formalism of the generalized gradient approximation~\cite{Perdew1996}.
Lastly, the effects of spin-polarization were found to be negligible for all systems after rigorous testing and screening and so were excluded from calculations in the dataset. 
For more detailed information and an in-depth analysis of some features of the DFT training set, see Ref.~\cite{startt2022compositional}.
}
\subsection*{Interatomic Potential Construction}

\newcommand{\br}{{\bf r}}
\newcommand{\bu}{{\bf u}}
\newcommand{\bU}{{\bf U}}
\newcommand{\bB}{{\bf B}}
\newcommand{\balpha}{{\boldsymbol\alpha}}
\newcommand{\bbeta}{{\boldsymbol\beta}}

The energy model is constructed using the standard form of SNAP that is described in more detail in earlier publications.\cite{Thompson2015, Wood2018}
The total potential energy of a configuration of atoms is written as the sum of atomic energies combined with an additional reference potential,
\begin{equation}
E({\bf r}^{N})=E_{ref}({\bf r}^{N})+\sum_{i=1}^{N}E_i({\bf r}^{N}),
\label{snapE}
\end{equation}
where $E$ is the total potential, ${\bf r}^{N}$ are the positions of the $N$ atoms in the configuration, $E_{ref}$ is the reference energy, and $E_i$ is the atomic energy of atom $i$.  The atomic energy of atom $i$ is expressed as a sum of the bispectrum components ${\bf B}_i$ for that atom weighted by regression coefficients 
\begin{eqnarray}
E_i({\bf r}^{N}) & = & 
{\bbeta}_{\nu_i} \cdot ({\bB}_i -  {\bB}_{0\nu_i}) \,,
\label{eqn:e}
\end{eqnarray}
where the elements of the vector ${\bbeta}_{\nu}$ are constant linear coefficients for atoms of element ${\nu}$ whose values are determined in training. 
The vector ${\bB}_i$ is a flattened list of bispectrum components
for atom $i$, while ${\bB}_{0\nu}$ is the list of bispectrum components
for an isolated atom of type $\nu$.  By construction, the energy of an isolated atom is zero.
The bispectrum components are real, rotationally invariant triple-products of four-dimensional hyperspherical harmonics $\bU_j$ \cite{Bartok2010}
    \begin{eqnarray}
        \label{eqn:b}
        B_{j_1j_2j}  &=& \bU_{j_1} \otimes_{j_1j_2}^j \bU_{j_2} \colon \bU_j^* \,, \end{eqnarray}
where symbol $\otimes_{j_1j_2}^j$ indicates a Clebsch-Gordan product of two matrices of arbitrary rank, while $:$ corresponds to an element-wise scalar product of two matrices of equal rank. The total hyperspherical harmonics for a central atom $i$ are written as sums over neighbor contributions,    
    \begin{eqnarray}
        \label{eqn:u}
        \bU_j &=& \bu_j({\bf 0}) + \sum_{k \in \mathcal{N}(i)}~f_c(r_{ik}) w_{\nu_k} \bu_j(\br_{ik}) \,,
    \end{eqnarray}
where the summation is over all neighbor atoms $k$ within a cutoff distance $R_{\nu_i \nu_k}$ of atom $i$.  Atoms of different chemical elements are distinguished by the element weights $w_{\nu}$.  The radial cutoff function $f_c(r)$ ensures that atomic contributions go smoothly to zero as $r$ approaches $R$ from below.  We used the 55 lowest order bispectrum components $B_{j_1j_2j}$ with half-integral indices restricted to the range $0 \le j_2 \le j_1 \le j \le 4$. The SNAP element weights and cutoff distances were optimized for each element using the genetic algorithm search described below.  To ensure strong repulsion at short separations between all pairs of atoms, a short-ranged ZBL\cite{zbl} reference potential was added ($Z$ = 44.5, $R_{cut}$ = 5.0~\AA).

\subsection*{Multi-compositional refractory CCA ML-IAP for MoNbTaTi}
{
In this section, we will briefly describe the fitting process and the performance of the MoNbTaTi refractory CCA ML-IAP. 
The primary target in fitting this potential was to ensure accuracy across a wide range of global composition space, with a particular focus on replicating DFT predicted elastic properties. 
Because elasticity is a property measured over an entire simulation cell, and not locally, we did not apply the local composition analysis techniques outlined above to optimize the fitting process. 
Instead, we followed the global composition sampling scheme detailed in the "Density functional theory calculations for training set" described above.
}

{
For the present the case of the MoNbTaTi CCA ML-IAP, two methodologies were combined: linear regression to generate individual IAPs and a genetic algorithm to explore parameter space. 
The first method is the core model connecting the DFT training set and a resulting IAP.
We use a linear regression scheme whose general form is: 
\begin{equation}
\hat{\boldsymbol\sigma} = \underset{\boldsymbol\beta}{\operatorname{argmin}} (\|\boldsymbol\epsilon \circ (D \boldsymbol\beta- T) \|^{2}-\gamma_{n}~\|\boldsymbol\beta\|^{n})
\label{euclid}
\end{equation}
\\
$\hat{\boldsymbol\sigma}$ represents regression error between the bispectrum descriptor predictions, $D$ and the training data reference energies, $T$. $\gamma_{n}~\|\boldsymbol\beta\|^{n}$ is a regularization penalty of order $n$ and weight $\gamma_{n}$. 
}

{
To implement this scheme, descriptors of the training structures from the DFT must be reproduced in MD simulations. 
For this purpose, we used the open-source software FitSNAP~\cite{Rohskopf2023} (GitHub link: \url{https://github.com/FitSNAP/FitSNAP}), which parses configurations from DFT into LAMMPS. 
Once each structures' corresponding bispectrum descriptors, reference energies, and reference forces have been calculated, FitSNAP retrieves that information from LAMMPS to form the D matrix from Eq. \ref{euclid}. 
The generation process is completed by solving for $\hat{\boldsymbol\beta}$  using singular value decomposition and outputting an IAP.
Though the first process as described above will generate an IAP for use in LAMMPS, it does not guarantee that it will be optimal for applications of interest.
To address the issue of tuning variables, we wrap a single-objective genetic algorithm (GA) around FitSNAP, as implemented by the DAKOTA optimization software~\cite{dakota}. 
The purpose of the GA is to take a user-generated selection of target IAP properties and discover the optimal set of variables to use within the core fitting calcuations in FitSNAP. 
To accomplish this, the user creates a series of short simulations in LAMMPS that are run and are used to evaluate a generated IAP's overall quality. 
}

{
Including multi-compositional data in the training set is not enough to guarantee that an IAP will perform well across those compositions. 
We found the key technique to be the setup of the simulations that feed into the GA's objective functions. 
For the purposes of this work, we aimed to optimize ML-IAPs on the first-principles elasticity data (see ``Density functional theory calculations for training set'' subsection above), which indicate that MoNbTaTi alloys undergo linear changes in elastic moduli with the enrichment and depletion of single elements. 
Thus, our post-fit testing and matching GA objective functions were designed to make the GA especially sensitive to the rate of change of the bulk ($B$) and shear ($G$) moduli as compositions are varied. 
Fitting the moduli to slopes instead of single-composition values forces the GA to favor not only low regression errors on single-composition fits, but also couple those errors' tendencies to favor accurate calculations of $B$ and $G$. 
To fit these slopes, it is also necessary to create objective functions that minimize errors on the components of the elasticity tensor, $\mathbb{C}_{xx}$. 
As the stable MoNbTaTi single-phase refractory CCAs take on BCC structure, only three of the tensor components ($\mathbb{C}_{11}$, $\mathbb{C}_{12}$, and $\mathbb{C}_{44}$), which we will also refer to as the elastic constants, need to be calculated per IAP and composition tested. 
Having obtained those, the calculation of the bulk and shear moduli is as follows:

\begin{equation}
B = \frac{\mathbb{C}_{11}+2\mathbb{C}_{12}}{3}
\label{bulk_mod}
\end{equation} \begin{equation}
G = \frac{1}{2} \left [ \frac{\mathbb{C}_{11}-\mathbb{C}_{12}+3\mathbb{C}_{44}}{5} + \frac{5\mathbb{C}_{44} \left ( \mathbb{C}_{11}-\mathbb{C}_{12} \right ) }{ 4\mathbb{C}_{44}+3\left (\mathbb{C}_{11}-\mathbb{C}_{12} \right ) }  \right ]
\label{shear_mod}
\end{equation} 

%

Once calculated for three separate compositions, a slope for $B$ and $G$ for the ML-IAP can be fit and tested against slopes calculated from the training data. 
}


\subsection*{Vacancy energy simulations}
{ 
In alloys, the vacancy formation energy of a given atomic species $\mu$, $E_{\rm vac}^{\mu}$, is found by first calculating the average cohesive energy of all atoms in a cell $E_{\rm cohesive}^{\rm all}$, removing an atom of one species $\mu$, and then recalculating the cell's new energy $E_{\rm removed}^{\mu}$. The difference of the new energy and the total potential energy corresponding to a cell one atom smaller gives the penalty for vacancy formation for that species $\mu$:

\begin{equation}
E_{\rm vac,f}^{\mu} = E_{\rm removed}^{\mu} - E_{\rm total}^{\rm all}\cdot\frac{N_{\rm cell}-1}{N_{\rm cell}}
\label{E_vac}
\end{equation}
}


\subsection*{Hybrid Monte Carlo/molecular dynamics (MC/MD) simulations}

{ 
To induce chemical reordering in the MoNbTaTi SNAP potential, we used a hybrid Monte Carlo/molecular dynamics procedure on one equiatomic random solid solution cube with side length of $s \cdot a_0=40 \cdot 3.2546$ \AA~, totalling 128,000 atoms. 
The random solid solution was initially relaxed at T = 300K for 10 ps in the NPT ensemble (Nose-Hoover barostat), using a 1 fs timestep.
For the hybrid algorthim, the same settings were used for all further MD steps.
The MC element type swaps were conducted at intervals of 50 MD steps (0.050 ps) using a series of \emph{fix atom/swap} in LAMMPS at a MC temperature of 300K.
In total, 150 ps of hybrid MC/MD (75,000 MC steps with 768,000 atom swap attempts, 138,748 accepted) were conducted to achieve the results found in Fig\@.~\ref{fig:fig5}a.
Warren-Cowley parameters~\cite{cowley1950approximate} were used to calculate trends in pair ordering for the 1st nearest-neighbor shell of central atom $i$ relative to elements A and B in Fig\@.~\ref{fig:fig5}b:
\begin{equation}
\alpha_{i}^{AB} = 1 - \frac{P_{i}^{AB}}{c_{j}}
\label{WCparam}
\end{equation}
Details of the convergence of potential energy through MC steps can be found in the Supplemental Information. 
}


\section*{DATA AVAILABILITY}

The full training data sets as well as all validation and test cases are available from the corresponding author upon reasonable request.





\section*{ACKNOWLEDGMENTS}
\label{sec:Acknowledgments}
{
This work was supported by the U.S. Department of Energy, Office of Fusion Energy Sciences (OFES) under Field Work Proposal Number 20-023149, and the the Center for Integrated Nanotechnologies, an Office of Science user facility operated for the U.S. Department of Energy.
This article has been authored by an employee of National Technology \& Engineering Solutions of Sandia, LLC under Contract No. DE-NA0003525 with the U.S. Department of Energy (DOE). The employee owns all right, title and interest in and to the article and is solely responsible for its contents. The United States Government retains and the publisher, by accepting the article for publication, acknowledges that the United States Government retains a non-exclusive, paid-up, irrevocable, world-wide license to publish or reproduce the published form of this article or allow others to do so, for United States Government purposes. The DOE will provide public access to these results of federally sponsored research in accordance with the DOE Public Access Plan https://www.energy.gov/downloads/doe-public-access-plan.
}

\section*{AUTHOR CONTRIBUTIONS}
\label{sec:Contribution}
{
MJM : Development of interatomic potential, performed MD simulations, implementation of local composition analysis code, chemical complexity analysis. 
JS : Performed DFT simulations, chemical complexity analysis.
APT : Implemented interatomic potential in LAMMPS, chemical complexity analysis.
RD : Validation of DFT, MD and interatomic potential.  
MAW : Development of interatomic potential, chemical complexity analysis.
All authors participated in conceiving the research and writing the manuscript.
}

\section*{COMPETING INTERESTS}

    The authors declare no competing interests.


\section*{ADDITIONAL INFORMATION}
{\bf Supplementary Information} is available for this paper.

\bibliographystyle{naturemag}  
\def\bibsection{\section*{\refname}}
\bibliography{bibliography.bib}

\begin{thebibliography}{10}
\expandafter\ifx\csname url\endcsname\relax
  \def\url#1{\texttt{#1}}\fi
\expandafter\ifx\csname urlprefix\endcsname\relax\def\urlprefix{URL }\fi
\providecommand{\bibinfo}[2]{#2}
\providecommand{\eprint}[2][]{\url{#2}}

\bibitem{senkov2015accelerated}
\bibinfo{author}{Senkov, O.}, \bibinfo{author}{Miller, J.},
  \bibinfo{author}{Miracle, D.} \& \bibinfo{author}{Woodward, C.}
\newblock \bibinfo{title}{Accelerated exploration of multi-principal element
  alloys for structural applications}.
\newblock \emph{\bibinfo{journal}{Calphad}} \textbf{\bibinfo{volume}{50}},
  \bibinfo{pages}{32--48} (\bibinfo{year}{2015}).
\newblock \urlprefix\url{https://doi.org/10.1016/j.calphad.2015.04.009}.

\bibitem{yao2016}
\bibinfo{author}{Yao, H.} \emph{et~al.}
\newblock \bibinfo{title}{{MoNbTaV} medium-entropy alloy}.
\newblock \emph{\bibinfo{journal}{Entropy}} \textbf{\bibinfo{volume}{18}},
  \bibinfo{pages}{189} (\bibinfo{year}{2016}).
\newblock \urlprefix\url{https://doi.org/10.3390/e18050189}.

\bibitem{senkov2018development}
\bibinfo{author}{Senkov, O.}, \bibinfo{author}{Miracle, D.},
  \bibinfo{author}{Chaput, K.} \& \bibinfo{author}{Couzini\'e, J.-P.}
\newblock \bibinfo{title}{Development and exploration of refractory high
  entropy alloys -- {A} review}.
\newblock \emph{\bibinfo{journal}{J. Mater. Res.}}
  \textbf{\bibinfo{volume}{33}}, \bibinfo{pages}{3092--3128}
  (\bibinfo{year}{2018}).
\newblock \urlprefix\url{https://doi.org/10.1557/jmr.2018.153}.

\bibitem{coury2019}
\bibinfo{author}{Coury, F.}, \bibinfo{author}{Kaufman, M.} \&
  \bibinfo{author}{Clarke, A.}
\newblock \bibinfo{title}{Solid-solution strengthening in refractory high
  entropy alloys}.
\newblock \emph{\bibinfo{journal}{Acta Mater.}} \textbf{\bibinfo{volume}{175}},
  \bibinfo{pages}{66--81} (\bibinfo{year}{2019}).
\newblock \urlprefix\url{https://doi.org/10.1016/j.actamat.2019.06.006}.

\bibitem{senkov2019high}
\bibinfo{author}{Senkov, O.}, \bibinfo{author}{Gorsse, S.} \&
  \bibinfo{author}{Miracle, D.}
\newblock \bibinfo{title}{High temperature strength of refractory complex
  concentrated alloys}.
\newblock \emph{\bibinfo{journal}{Acta Mater.}} \textbf{\bibinfo{volume}{175}},
  \bibinfo{pages}{394--405} (\bibinfo{year}{2019}).
\newblock \urlprefix\url{https://doi.org/10.1016/j.actamat.2019.06.032}.

\bibitem{startt2022compositional}
\bibinfo{author}{Startt, J.}, \bibinfo{author}{Kustas, A.},
  \bibinfo{author}{Pegues, J.}, \bibinfo{author}{Yang, P.} \&
  \bibinfo{author}{Dingreville, R.}
\newblock \bibinfo{title}{Compositional effects on the mechanical and thermal
  properties of {MoNbTaTi} refractory complex concentrated alloys}.
\newblock \emph{\bibinfo{journal}{Mater. Des.}} \textbf{\bibinfo{volume}{213}},
  \bibinfo{pages}{110311} (\bibinfo{year}{2022}).
\newblock \urlprefix\url{https://doi.org/10.1016/j.matdes.2021.110311}.

\bibitem{cui2020wear}
\bibinfo{author}{Cui, Y.}, \bibinfo{author}{Shen, J.},
  \bibinfo{author}{Manladan, S.~M.}, \bibinfo{author}{Geng, K.} \&
  \bibinfo{author}{Hu, S.}
\newblock \bibinfo{title}{Wear resistance of {FeCoCrNiMnAl$_\textrm{x}$}
  high-entropy alloy coatings at high temperature}.
\newblock \emph{\bibinfo{journal}{Appl. Surf. Sci.}}
  \textbf{\bibinfo{volume}{512}}, \bibinfo{pages}{145736}
  (\bibinfo{year}{2020}).
\newblock \urlprefix\url{https://doi.org/10.1016/j.apsusc.2020.145736}.

\bibitem{du2021mechanical}
\bibinfo{author}{Du, Y.} \emph{et~al.}
\newblock \bibinfo{title}{Mechanical and tribological performance of
  {CoCrNiHf$_\textrm{x}$} eutectic medium-entropy alloys}.
\newblock \emph{\bibinfo{journal}{J. Mater. Sci. Technol.}}
  \textbf{\bibinfo{volume}{90}}, \bibinfo{pages}{194--204}
  (\bibinfo{year}{2021}).
\newblock \urlprefix\url{https://doi.org/10.1016/j.jmst.2021.03.023}.

\bibitem{liu2019effect}
\bibinfo{author}{Liu, Y.} \emph{et~al.}
\newblock \bibinfo{title}{Effect of {Al} content on high temperature oxidation
  resistance of {Al$_{\textrm x}$CoCrCuFeNi} high entropy alloys (x= 0, 0.5, 1,
  1.5, 2)}.
\newblock \emph{\bibinfo{journal}{Vacuum}} \textbf{\bibinfo{volume}{169}},
  \bibinfo{pages}{108837} (\bibinfo{year}{2019}).
\newblock \urlprefix\url{https://doi.org/10.1016/j.vacuum.2019.108837}.

\bibitem{esmaily2020high}
\bibinfo{author}{Esmaily, M.} \emph{et~al.}
\newblock \bibinfo{title}{High-temperature oxidation behaviour of {Al$_{\textrm
  x}$FeCrCoNi} and {AlTiVCr} compositionally complex alloys}.
\newblock \emph{\bibinfo{journal}{npj Mater. Degrad.}}
  \textbf{\bibinfo{volume}{4}}, \bibinfo{pages}{1--10} (\bibinfo{year}{2020}).
\newblock \urlprefix\url{https://doi.org/10.1038/s41529-020-00129-2}.

\bibitem{scully2020controlling}
\bibinfo{author}{Scully, J.~R.} \emph{et~al.}
\newblock \bibinfo{title}{Controlling the corrosion resistance of
  multi-principal element alloys}.
\newblock \emph{\bibinfo{journal}{Scripta Mater.}}
  \textbf{\bibinfo{volume}{188}}, \bibinfo{pages}{96--101}
  (\bibinfo{year}{2020}).
\newblock \urlprefix\url{https://doi.org/10.1016/j.scriptamat.2020.06.065}.

\bibitem{zhao2018effect}
\bibinfo{author}{Zhao, R.-F.}, \bibinfo{author}{Ren, B.},
  \bibinfo{author}{Zhang, G.-P.}, \bibinfo{author}{Liu, Z.-X.} \&
  \bibinfo{author}{Zhang, J.-J.}
\newblock \bibinfo{title}{Effect of {Co} content on the phase transition and
  magnetic properties of {Co$_\textrm{x}$CrCuFeMnNi} high-entropy alloy
  powders}.
\newblock \emph{\bibinfo{journal}{J. Magn. Magn. Mater.}}
  \textbf{\bibinfo{volume}{468}}, \bibinfo{pages}{14--24}
  (\bibinfo{year}{2018}).
\newblock \urlprefix\url{https://doi.org/10.1016/j.jmmm.2018.07.072}.

\bibitem{chaudhary2021accelerated}
\bibinfo{author}{Chaudhary, V.}, \bibinfo{author}{Chaudhary, R.},
  \bibinfo{author}{Banerjee, R.} \& \bibinfo{author}{Ramanujan, R.}
\newblock \bibinfo{title}{Accelerated and conventional development of magnetic
  high entropy alloys}.
\newblock \emph{\bibinfo{journal}{Mater. Today}} \textbf{\bibinfo{volume}{49}},
  \bibinfo{pages}{231--252} (\bibinfo{year}{2021}).
\newblock \urlprefix\url{https://doi.org/10.1016/j.mattod.2021.03.018}.

\bibitem{mishra2021design}
\bibinfo{author}{Mishra, R.~K.}, \bibinfo{author}{Kumari, P.},
  \bibinfo{author}{Gupta, A.~K.} \& \bibinfo{author}{Shahi, R.~R.}
\newblock \bibinfo{title}{Design and development of
  {Co35Cr5Fe$_{20-\textrm{x}}$Ni$_{20+\textrm{x}}$Ti$_20$} high entropy alloy
  with excellent magnetic softness}.
\newblock \emph{\bibinfo{journal}{J. Alloys Compd}}
  \textbf{\bibinfo{volume}{889}}, \bibinfo{pages}{161773}
  (\bibinfo{year}{2021}).
\newblock \urlprefix\url{https://doi.org/10.1016/j.jallcom.2021.161773}.

\bibitem{byggmastar2021modeling}
\bibinfo{author}{Byggm{\"a}star, J.}, \bibinfo{author}{Nordlund, K.} \&
  \bibinfo{author}{Djurabekova, F.}
\newblock \bibinfo{title}{Modeling refractory high-entropy alloys with
  efficient machine-learned interatomic potentials: {D}efects and segregation}.
\newblock \emph{\bibinfo{journal}{Phys. Rev. B}}
  \textbf{\bibinfo{volume}{104}}, \bibinfo{pages}{104101}
  (\bibinfo{year}{2021}).
\newblock \urlprefix\url{https://doi.org/10.1103/PhysRevB.104.104101}.

\bibitem{li2020complex}
\bibinfo{author}{Li, X.-G.}, \bibinfo{author}{Chen, C.},
  \bibinfo{author}{Zheng, H.}, \bibinfo{author}{Zuo, Y.} \&
  \bibinfo{author}{Ong, S.~P.}
\newblock \bibinfo{title}{Complex strengthening mechanisms in the {NbMoTaW}
  multi-principal element alloy}.
\newblock \emph{\bibinfo{journal}{npj Comput. Mater.}}
  \textbf{\bibinfo{volume}{6}}, \bibinfo{pages}{1--10} (\bibinfo{year}{2020}).
\newblock \urlprefix\url{https://doi.org/10.1038/s41524-020-0339-0}.

\bibitem{yin2021atomistic}
\bibinfo{author}{Yin, S.} \emph{et~al.}
\newblock \bibinfo{title}{Atomistic simulations of dislocation mobility in
  refractory high-entropy alloys and the effect of chemical short-range order}.
\newblock \emph{\bibinfo{journal}{Nat. Commun.}} \textbf{\bibinfo{volume}{12}},
  \bibinfo{pages}{1--14} (\bibinfo{year}{2021}).
\newblock \urlprefix\url{https://doi.org/10.1038/s41467-021-25134-0}.

\bibitem{Rohskopf2023}
\bibinfo{author}{Rohskopf, A.} \emph{et~al.}
\newblock \bibinfo{title}{Fitsnap: Atomistic machine learning with lammps}.
\newblock \emph{\bibinfo{journal}{Journal of Open Source Software}}
  \textbf{\bibinfo{volume}{8}}, \bibinfo{pages}{5118} (\bibinfo{year}{2023}).
\newblock \urlprefix\url{https://doi.org/10.21105/joss.05118}.

\bibitem{dakota}
\bibinfo{author}{Adams, B.} \emph{et~al.}
\newblock \bibinfo{title}{Dakota: a multilevel parallel object-oriented
  framework for design optimization, parameter estimation, uncertainty
  quantification, and sensitivity analysis}.
\newblock \bibinfo{type}{Tech. Rep.} \bibinfo{number}{SAND2010-2183},
  \bibinfo{institution}{Sandia National Laboratories} (\bibinfo{year}{2009}).
\newblock \urlprefix\url{https://doi.org/10.2172/1630693}.

\bibitem{zhou2022vacancy}
\bibinfo{author}{Zhou, X.}, \bibinfo{author}{He, S.} \&
  \bibinfo{author}{Marian, J.}
\newblock \bibinfo{title}{Vacancy energetics and diffusivities in the
  equiatomic multielement {Nb-Mo-Ta-W} alloy}.
\newblock \emph{\bibinfo{journal}{Materials}} \textbf{\bibinfo{volume}{15}},
  \bibinfo{pages}{5468} (\bibinfo{year}{2022}).
\newblock \urlprefix\url{https://doi.org/10.3390/ma15155468}.

\bibitem{cusentino2020compositional}
\bibinfo{author}{Cusentino, M.}, \bibinfo{author}{Wood, M.} \&
  \bibinfo{author}{Dingreville, R.}
\newblock \bibinfo{title}{Compositional and structural origins of radiation
  damage mitigation in high-entropy alloys}.
\newblock \emph{\bibinfo{journal}{J. Appl. Phys.}}
  \textbf{\bibinfo{volume}{128}}, \bibinfo{pages}{125904}
  (\bibinfo{year}{2020}).
\newblock \urlprefix\url{https://doi.org/10.1063/5.0024014}.

\bibitem{li2019first}
\bibinfo{author}{Li, C.} \emph{et~al.}
\newblock \bibinfo{title}{First principle study of magnetism and vacancy
  energetics in a near equimolar {NiFeMnCr} high entropy alloy}.
\newblock \emph{\bibinfo{journal}{J. Appl. Phys.}}
  \textbf{\bibinfo{volume}{125}}, \bibinfo{pages}{155103}
  (\bibinfo{year}{2019}).
\newblock \urlprefix\url{https://doi.org/10.1063/1.5086172}.

\bibitem{stukowski2012elastic}
\bibinfo{author}{Stukowski, A.} \& \bibinfo{author}{Arsenlis, A.}
\newblock \bibinfo{title}{On the elastic--plastic decomposition of crystal
  deformation at the atomic scale}.
\newblock \emph{\bibinfo{journal}{Model. Simul. Mat. Sci. Eng.}}
  \textbf{\bibinfo{volume}{20}}, \bibinfo{pages}{035012}
  (\bibinfo{year}{2012}).
\newblock \urlprefix\url{https://doi.org/10.1088/0965-0393/20/3/035012}.

\bibitem{ferrari2020frontiers}
\bibinfo{author}{Ferrari, A.} \emph{et~al.}
\newblock \bibinfo{title}{Frontiers in atomistic simulations of high entropy
  alloys}.
\newblock \emph{\bibinfo{journal}{J. Appl. Phys.}}
  \textbf{\bibinfo{volume}{128}}, \bibinfo{pages}{150901}
  (\bibinfo{year}{2020}).
\newblock \urlprefix\url{https://doi.org/10.1063/5.0025310}.

\bibitem{antillon2021}
\bibinfo{author}{Antillon, E.}, \bibinfo{author}{Woodward, C.},
  \bibinfo{author}{Rao, S.~I.} \& \bibinfo{author}{Akdim, B.}
\newblock \bibinfo{title}{Chemical short range order strengthening in {BCC}
  complex concentrated alloys}.
\newblock \emph{\bibinfo{journal}{Acta Mater.}} \textbf{\bibinfo{volume}{215}},
  \bibinfo{pages}{117012} (\bibinfo{year}{2021}).
\newblock \urlprefix\url{https://doi.org/10.1016/j.actamat.2021.117012}.

\bibitem{miracle2017critical}
\bibinfo{author}{Miracle, D.~B.} \& \bibinfo{author}{Senkov, O.~N.}
\newblock \bibinfo{title}{A critical review of high entropy alloys and related
  concepts}.
\newblock \emph{\bibinfo{journal}{Acta Mater.}} \textbf{\bibinfo{volume}{122}},
  \bibinfo{pages}{448--511} (\bibinfo{year}{2017}).
\newblock \urlprefix\url{https://doi.org/10.1016/j.actamat.2016.08.081}.

\bibitem{widom2014hybrid}
\bibinfo{author}{Widom, M.}, \bibinfo{author}{Huhn, W.~P.},
  \bibinfo{author}{Maiti, S.} \& \bibinfo{author}{Steurer, W.}
\newblock \bibinfo{title}{Hybrid {Monte Carlo}/molecular dynamics simulation of
  a refractory metal high entropy alloy}.
\newblock \emph{\bibinfo{journal}{Metall. Mater. Trans. A}}
  \textbf{\bibinfo{volume}{45}}, \bibinfo{pages}{196--200}
  (\bibinfo{year}{2014}).
\newblock \urlprefix\url{https://doi.org/10.1038/s41524-020-0339-0}.

\bibitem{kormann2017long}
\bibinfo{author}{K{\"o}rmann, F.}, \bibinfo{author}{Ruban, A.~V.} \&
  \bibinfo{author}{Sluiter, M.~H.}
\newblock \bibinfo{title}{Long-ranged interactions in bcc {NbMoTaW}
  high-entropy alloys}.
\newblock \emph{\bibinfo{journal}{Mater. Res. Lett.}}
  \textbf{\bibinfo{volume}{5}}, \bibinfo{pages}{35--40} (\bibinfo{year}{2017}).
\newblock \urlprefix\url{https://doi.org/10.1080/21663831.2016.1198837}.

\bibitem{cowley1950approximate}
\bibinfo{author}{Cowley, J.~M.}
\newblock \bibinfo{title}{An approximate theory of order in alloys}.
\newblock \emph{\bibinfo{journal}{Phys. Rev.}} \textbf{\bibinfo{volume}{77}},
  \bibinfo{pages}{669} (\bibinfo{year}{1950}).
\newblock \urlprefix\url{https://doi.org/10.1103/PhysRev.77.669}.

\bibitem{zhang2020short}
\bibinfo{author}{Zhang, R.} \emph{et~al.}
\newblock \bibinfo{title}{Short-range order and its impact on the {CrCoNi}
  medium-entropy alloy}.
\newblock \emph{\bibinfo{journal}{Nature}} \textbf{\bibinfo{volume}{581}},
  \bibinfo{pages}{283--287} (\bibinfo{year}{2020}).
\newblock \urlprefix\url{https://doi.org/10.1038/s41586-020-2275-z}.

\bibitem{Kresse1993}
\bibinfo{author}{Kresse, G.} \& \bibinfo{author}{Hafner, J.}
\newblock \bibinfo{title}{{Ab initio molecular dynamics for liquid metals}}.
\newblock \emph{\bibinfo{journal}{Phys. Rev. B}} \textbf{\bibinfo{volume}{47}},
  \bibinfo{pages}{558--561} (\bibinfo{year}{1993}).
\newblock \urlprefix\url{https://doi.org/10.1103/PhysRevB.47.558}.

\bibitem{Kresse1994}
\bibinfo{author}{Kresse, G.} \& \bibinfo{author}{Hafner, J.}
\newblock \bibinfo{title}{{Ab initio molecular-dynamics simulation of the
  liquid-metalamorphous- semiconductor transition in germanium}}.
\newblock \emph{\bibinfo{journal}{Phys. Rev. B}} \textbf{\bibinfo{volume}{49}},
  \bibinfo{pages}{14251--14269} (\bibinfo{year}{1994}).
\newblock \urlprefix\url{https://doi.org/10.1103/PhysRevB.49.14251}.

\bibitem{Kresse1995}
\bibinfo{author}{Kresse, G.}
\newblock \bibinfo{title}{{Ab initio molecular dynamics for liquid metals}}.
\newblock \emph{\bibinfo{journal}{J. Non-Cryst. Solids}}
  \textbf{\bibinfo{volume}{192-193}}, \bibinfo{pages}{222--229}
  (\bibinfo{year}{1995}).
\newblock \urlprefix\url{https://doi.org/10.1016/0022-3093(95)00355-X}.

\bibitem{Blochl1994}
\bibinfo{author}{Blochl, P.~E.}
\newblock \bibinfo{title}{{Projector augmented-wave method}}.
\newblock \emph{\bibinfo{journal}{Phys. Rev. B}} \textbf{\bibinfo{volume}{50}},
  \bibinfo{pages}{17979} (\bibinfo{year}{1994}).
\newblock \urlprefix\url{https://doi.org/10.1103/PhysRevB.50.17953}.

\bibitem{Kresse1999}
\bibinfo{author}{Kresse, G.} \& \bibinfo{author}{Joubert, D.}
\newblock \bibinfo{title}{{From ultrasoft pseudopotentials to the projector
  augmented-wave method}}.
\newblock \emph{\bibinfo{journal}{Phys. Rev. B}} \textbf{\bibinfo{volume}{59}},
  \bibinfo{pages}{1758--1775} (\bibinfo{year}{1999}).
\newblock \urlprefix\url{https://doi.org/10.1103/PhysRevB.59.1758}.

\bibitem{Vandewall2013}
\bibinfo{author}{Van~de Walle, A.} \emph{et~al.}
\newblock \bibinfo{title}{Efficient stochastic generation of special
  quasirandom structures}.
\newblock \emph{\bibinfo{journal}{Calphad}} \textbf{\bibinfo{volume}{42}},
  \bibinfo{pages}{13--18} (\bibinfo{year}{2013}).
\newblock \urlprefix\url{https://doi.org/10.1016/j.calphad.2013.06.006}.

\bibitem{Anubhav2013}
\bibinfo{author}{Jain, A.} \emph{et~al.}
\newblock \bibinfo{title}{Commentary: The materials project: A materials genome
  approach to accelerating materials innovation}.
\newblock \emph{\bibinfo{journal}{APL Mater.}} \textbf{\bibinfo{volume}{1}},
  \bibinfo{pages}{011002} (\bibinfo{year}{2013}).
\newblock \urlprefix\url{https://doi.org/10.1063/1.4812323}.

\bibitem{Perdew1996}
\bibinfo{author}{Perdew, J.~P.}, \bibinfo{author}{Burke, K.} \&
  \bibinfo{author}{Ernzerhof, M.}
\newblock \bibinfo{title}{{Generalized gradient approximation made simple}}.
\newblock \emph{\bibinfo{journal}{Phys. Rev. Lett.}}
  \textbf{\bibinfo{volume}{77}}, \bibinfo{pages}{3865--3868}
  (\bibinfo{year}{1996}).
\newblock \urlprefix\url{https://doi.org/10.1103/PhysRevLett.77.3865}.

\bibitem{Thompson2015}
\bibinfo{author}{Thompson, A.~P.}, \bibinfo{author}{Swiler, L.~P.},
  \bibinfo{author}{Trott, C.~R.}, \bibinfo{author}{Foiles, S.~M.} \&
  \bibinfo{author}{Tucker, G.~J.}
\newblock \bibinfo{title}{{Spectral neighbor analysis method for automated
  generation of quantum-accurate interatomic potentials}}.
\newblock \emph{\bibinfo{journal}{J. Comput. Phys.}}
  \textbf{\bibinfo{volume}{285}}, \bibinfo{pages}{316--330}
  (\bibinfo{year}{2015}).
\newblock \urlprefix\url{http://dx.doi.org/10.1016/j.jcp.2014.12.018}.

\bibitem{Wood2018}
\bibinfo{author}{Wood, M.~A.} \& \bibinfo{author}{Thompson, A.~P.}
\newblock \bibinfo{title}{{Extending the accuracy of the SNAP interatomic
  potential form}}.
\newblock \emph{\bibinfo{journal}{J. Chem. Phys.}}
  \textbf{\bibinfo{volume}{148}}, \bibinfo{pages}{1--10}
  (\bibinfo{year}{2018}).
\newblock \urlprefix\url{http://dx.doi.org/10.1063/1.5017641}.
\newblock \eprint{1711.11131}.

\bibitem{Bartok2010}
\bibinfo{author}{Bart{\'o}k, A.~P.}, \bibinfo{author}{Payne, M.~C.},
  \bibinfo{author}{Kondor, R.} \& \bibinfo{author}{Cs{\'a}nyi, G.}
\newblock \bibinfo{title}{Gaussian approximation potentials: The accuracy of
  quantum mechanics, without the electrons}.
\newblock \emph{\bibinfo{journal}{Phys. Rev. Lett.}}
  \textbf{\bibinfo{volume}{104}}, \bibinfo{pages}{136403}
  (\bibinfo{year}{2010}).
\newblock \urlprefix\url{https://doi.org/10.1103/PhysRevLett.104.136403}.

\bibitem{zbl}
\bibinfo{author}{Ziegler, J.}, \bibinfo{author}{Biersack, J.} \&
  \bibinfo{author}{Littmark, U.}
\newblock \emph{\bibinfo{title}{The Stopping and Range of Ions in Solids}}
  (\bibinfo{publisher}{Pergamon}, \bibinfo{year}{1985}).

\bibitem{ovito}
\bibinfo{author}{Stukowski, A.}
\newblock \bibinfo{title}{{Visualization and analysis of atomistic simulation
  data with {OVITO}-the Open Visualization Tool}}.
\newblock \emph{\bibinfo{journal}{{Model. Simul. Mat. Sci. Eng.}}}
  \textbf{\bibinfo{volume}{{18}}} (\bibinfo{year}{{2010}}).
\newblock \urlprefix\url{https://doi.org/10.1088/0965-0393/18/1/015012}.

\end{thebibliography}

\clearpage
\setcounter{figure}{0}
\renewcommand{\figurename}{SUPPLEMENTAL FIG.}


 \begin{figure*}[h!t]
        \centering
        \includegraphics[width=\textwidth]{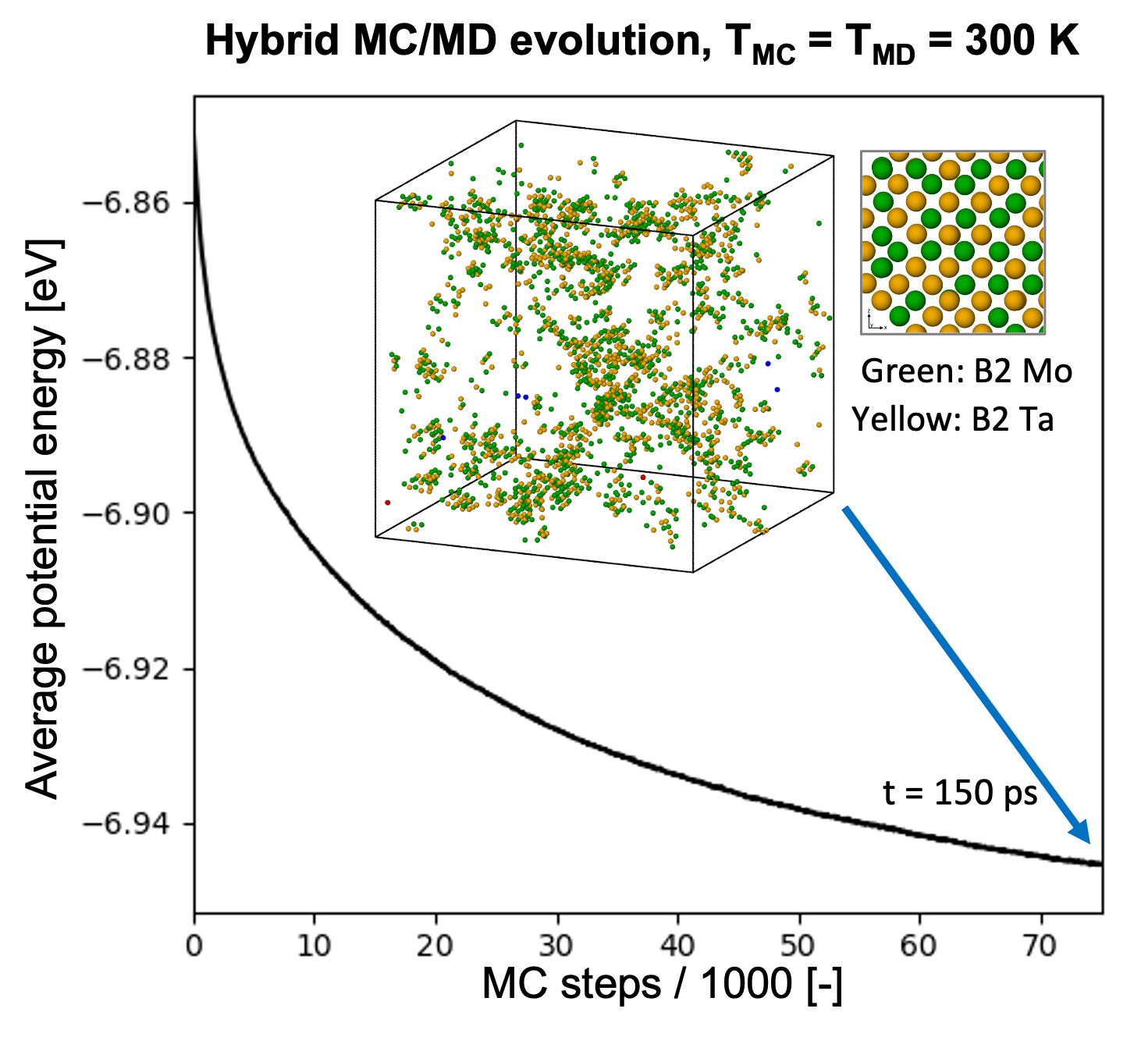}

        \caption{
        {\bf a} Average potential energy per atom vs. Monte Carlo (MC) steps, showing the reduction in overall system energy due to successful atom swaps at $T = 300 K$. The simulations were stopped at 150 ps (150,000 MD steps, 75,000 MC steps, 768,000 total swap attempts with 138,748 successful swaps). 
        The inset shows 2,235 Mo (green) and Ta (yellow) atoms identified as B2-ordered by the Ovito software's 'Polyhedral Template Matching' chemical ordering modifier \cite{ovito}.
        }
        \label{fig:sfig_MCMD}
 \end{figure*}


\end{document}